\documentclass[a4paper,reprint,twocolumn,amssymb,aps,prl,groupedaddress,superscriptaddress]{revtex4-2}
\usepackage{graphicx}
\usepackage{epsf}
\usepackage{bm}
\usepackage{amsmath}
\usepackage{amsfonts}
\usepackage{amssymb}
\usepackage{epstopdf}
\usepackage{natbib}
\usepackage{color}
\usepackage[dvipsnames]{xcolor}
\usepackage{verbatim}
\usepackage{multirow}
\usepackage{bm}
\usepackage{microtype}
\usepackage[colorlinks = true,
            linkcolor = blue,
            urlcolor  = blue,
            citecolor = blue,
            anchorcolor = blue]{hyperref}

\usepackage{amsmath}
\usepackage[capitalize]{cleveref}
\usepackage[normalem]{ulem}

\makeatletter\let\expandableinput\@@input\makeatother

\begin{document}

\title{$\Lambda_{\rm s}$CDM model: A promising scenario for alleviation of cosmological tensions}

\author{\"{O}zg\"{u}r Akarsu}
\email{akarsuo@itu.edu.tr}
\affiliation{Department of Physics, Istanbul Technical University, Maslak 34469 Istanbul, Turkey}

\author{Eleonora Di Valentino}
\email{e.divalentino@sheffield.ac.uk}
\affiliation{School of Mathematics and Statistics, University of Sheffield, Hounsfield Road, Sheffield S3 7RH, United Kingdom}

\author{Suresh Kumar}
\email{suresh.math@igu.ac.in}
\affiliation{Department of Mathematics, Indira Gandhi University, Meerpur, Haryana 122502, India}

\author{Rafael C. Nunes}
\email{rafadcnunes@gmail.com}
\affiliation{Instituto de F\'{i}sica, Universidade Federal do Rio Grande do Sul, 91501-970 Porto Alegre RS, Brazil}
\affiliation{Divis\~ao de Astrof\'isica, Instituto Nacional de Pesquisas Espaciais, Avenida dos Astronautas 1758, S\~ao Jos\'e dos Campos, 12227-010, SP, Brazil}

\author{J. Alberto Vazquez}
\email{javazquez@icf.unam.mx}
\affiliation{Instituto de Ciencias F\'isicas, Universidad Nacional Aut\'onoma de M\'exico, Cuernavaca, Morelos, 62210, M\'exico}

\author{Anita Yadav}
\email{anita.math.rs@igu.ac.in }
\affiliation{Department of Mathematics, Indira Gandhi University, Meerpur, Haryana 122502, India}

\begin{abstract}
We present a comprehensive analysis of the $\Lambda_{\rm s}$CDM model, which explores the recent conjecture suggesting a rapid transition of the Universe from anti-de Sitter vacua to de Sitter vacua (viz., the cosmological constant switches sign from negative to positive) at redshift ${z_\dagger\sim 2}$, inspired by the graduated dark energy (gDE) model. Our analysis shows that, predicting $z_\dagger\approx1.7$, $\Lambda_{\rm s}$CDM simultaneously addresses the major cosmological tensions of the standard $\Lambda$CDM model, viz., the Hubble constant $H_0$, the Type Ia Supernovae absolute magnitude $M_{\rm B}$, and the growth parameter $S_8$ tensions, along with other less significant tensions such as the BAO Lyman-$\alpha$ discrepancy. 
\end{abstract}


\maketitle
%
\textit{Introduction} -- The standard Lambda Cold Dark Matter ($\Lambda$CDM) scenario provides a wonderful fit for the majority of astrophysical 
and cosmological observations carried out over the past decades. However up to recently, in the new era of high-precision 
cosmology, some discrepancies became statistically significant while analyzing different data sets, placing it in a crossroad. 
This pivotal situation has compelled the scientific community to embark on a quest for alternative explanations, either 
rooted in novel physics or by the identification of potential systematic errors in the data. See~\cite{DiValentino:2021izs,Abdalla:2022yfr,Perivolaropoulos:2021jda,DiValentino:2022fjm} for recent reviews.

\vspace{-0.03cm}

The most statistically significant disagreement is in the value of the Hubble constant, $H_0$, between the  Planck-Cosmic Microwave Background (CMB)~\cite{Planck:2018vyg} estimate, assuming the standard $\Lambda$CDM model, and the direct local distance ladder measurements conducted by the SH0ES team~\cite{Riess:2021jrx, Riess:2022mme,Murakami:2023xuy}, reaching a significance of more than 5$\sigma$. Nevertheless the lower value of $H_0$ derived from the Planck-CMB data is in agreement with Baryon Acoustic Oscillations (BAO)+Big Bang Nucleosynthesis (BBN)~\cite{Schoneberg:2022ggi,Cuceu:2019for}, and also with other CMB experiments such as ACT-DR4 and SPT-3G~\cite{ACT:2020gnv,SPT-3G:2022hvq}. 
Conversely, the higher value of $H_0$ found by SH0ES based on the Supernovae calibrated by Cepheids is in agreement with all the local $H_0$ measurements~\cite{Ward:2022ghz,
CMB-S4:2016ple,  Sanders:2023jkl,Freedman:2021ahq,Anand:2021sum, Blakeslee:2021rqi,
deJaeger:2022lit}. Recent reanalysis of the Tip of the Red Giant Branch (TRGB) sample also shows that these local distance indicators are consistent with values of $H_0$ from Cepheids~\cite{Anderson:2023aga,Scolnic:2023mrv}.
Also, it has been argued that the $H_0$ tension is in fact a tension on the Type Ia Supernovae (SNIa) absolute magnitude 
$M_{\rm B}$~\cite{Efstathiou:2021ocp,Camarena:2021jlr,Nunes:2021zzi}, because the SH0ES $H_0$ measurement comes directly 
from $M_{\rm B}$ estimates. 

\vspace{-0.04cm}

On the other hand, within the framework of $\Lambda$CDM, the CMB measurements from Planck and ACT-DR4~\cite{Aiola_2020} 
indicate values of $S_8 = \sigma_8 \sqrt{\Omega_{\rm m}/0.3}$ that exhibit a 1.7$\sigma$--3$\sigma$ tension with 
those inferred from various weak lensing~\cite{DES:2021vln,DES:2021bvc,DES:2022ygi,KiDS:2020suj,Li:2023tui,Dalal:2023olq,Kilo-DegreeSurvey:2023gfr}, galaxy clustering, and redshift-space distortion measurements (see~\cite{DiValentino:2020vvd,Nunes:2021ipq, Perivolaropoulos:2021jda, Heymans:2020gsg, Abdalla:2022yfr, Sugiyama:2023}). While the status of the $S_8$ discrepancy may be somewhat 
less definitive compared to the $H_0$ tension, there is a clear trend of lower structure growth from low redshift Large Scale Structure probes compared to CMB.
Several other tensions and anomalies have 
recently emerged in the literature, e.g., BAO-Ly-$\alpha$ discrepancies, physical baryon density anomalies, age of the 
Universe issues, among others~\cite{Abdalla:2022yfr,Perivolaropoulos:2021jda}.  Although individually these tensions may not be very significant, when considered collectively, they could point to the existence of missing components in the current standard cosmological model. The search for a satisfactory 
explanation for these discrepancies, tensions and anomalies, either through systematic effects in data or 
new physics, has been a central theme in cosmology over the past few years.

\vspace{-0.04cm}

In this \textit{Letter}, we present an observational investigation of the sign-switching Lambda Cold Dark Matter ($\Lambda_{\rm s}$CDM) model, which takes into account the possibility that the Universe has recently, at redshift ${z\sim2}$, undergone a phase of rapid transition from anti-de Sitter (AdS) vacua to de Sitter (dS) vacua (viz., the cosmological constant switches sign)~\cite{Akarsu:2019hmw,Akarsu:2021fol,Akarsu:2022typ}. 
We show that this theory-framework can simultaneously resolve the major cosmological tensions,  
viz., the $H_0$, $M_{\rm B}$, and $S_8$ tensions, currently present in $\Lambda$CDM. If these tensions grow in significance with new and future data, and in the absence of clear systematic effects explaining their origins, 
we should look at $\Lambda_{\rm s}$CDM as a candidate that can direct us towards a new standard model of cosmology.

\textit{$\Lambda_{\rm s}$CDM model} -- The $\Lambda_{\rm s}$CDM model is inspired by the recent conjecture that the 
universe went through a spontaneous AdS-dS transition characterized by a sign-switching cosmological constant 
($\Lambda_{\rm s}$) at ${z\sim2}$~\cite{Akarsu:2019hmw,Akarsu:2021fol,Akarsu:2022typ}. This conjecture was proposed following the promising observational findings in the graduated dark energy (gDE) model, which showed that its density smoothly 
transitioning from negative to positive values rapidly enough at ${z\sim2}$ can simultaneously address the $H_0$ and BAO-Ly-$\alpha$ 
discrepancies~\cite{Akarsu:2019hmw}. The theoretical advantages of $\Lambda_{\rm s}$ 
over gDE further bolstered this conjecture~\cite{Akarsu:2021fol,Akarsu:2022typ}. The simplest 
$\Lambda_{\rm s}$CDM model was constructed phenomenologically by replacing the cosmological constant ($\Lambda$) 
of the standard $\Lambda$CDM model with an abrupt sign-switching cosmological constant ($\Lambda_{\rm s}$) at a 
redshift $z_\dagger$, which comes as the only additional free parameter. The present-day value of $\Lambda_{\rm s}$ 
is denoted as $\Lambda_{\rm s0}$, and the replacement is defined as:
\begin{equation}
    \Lambda\quad\rightarrow\quad\Lambda_{\rm s}\equiv\Lambda_{\rm s0}\,{\rm sgn}[z_\dagger-z],
    \label{eq:ssdeff}
\end{equation}
where the sign-switching transition of $\Lambda_{\rm s}$ is implemented by the signum function (sgn), which should be taken as an idealized description of a rapid transition phenomenon, such as a phase transition, from AdS vacua provided by $-\Lambda_{\rm s0}$ 
to dS vacua provided by $\Lambda_{\rm s0}$, or DE models such as gDE that can mimic this behavior~\cite{Akarsu:2022typ}.

The $\Lambda_{\rm s}$CDM model was first analysed using the Planck CMB data, followed by the inclusion of the full BAO data up to $z=2.36$ (viz., Ly-$\alpha$ DR14, BAO-Galaxy consensus, MGS and 6dFGS) in~\cite{Akarsu:2021fol}. It was shown that its consistency with CMB provides $H_0$ and $M_{\rm B}$ values that are inversely correlated with $z_\dagger$. 
Specifically, the analysis found that for $z_\dagger\sim1.6$, the model predicts values of $H_0\approx73.4~{\rm km\, s^{-1}\, Mpc^{-1}}$ and $M_B\approx-19.25\,{\rm mag}$, in  excellent agreement with the SH0ES measurements~\cite{Riess:2021jrx,Camarena:2021dcvm,Camarena:w2023rsd}. Additionally, provided that $z_\dagger\lesssim2.3$, it achieves an excellent fit to the Ly-$\alpha$ data. This was explained because $\Lambda_{\rm s}$CDM is exactly $\Lambda$CDM, except having $\Lambda_{\rm s}=-\Lambda_{\rm s0}<0$ for $z>z_\dagger$ (thereby respects the internal consistency of the SH0ES $H_0$ estimates utilizing $M_{\rm B}$ and leaves the standard pre-recombination universe untouched), and the fact that the comoving angular diameter distance to last scattering $D_M(z_*)=c\int_0^{z_{*}}H^{-1}{\rm d}{z}$ ($z_*\approx1090$) is strictly fixed by the CMB power spectra, a reduction in $H(z)$ for $z>z_{\dagger}$ compared to $\Lambda$CDM must be compensated by an enhancement in $H(z)$ for $z<z_\dagger$ resulting in higher $H_0$ and fainter $M_{\rm B}$. It turns out that, being consistent with Planck CMB for $z_\dagger\gtrsim1.5$, $\Lambda_{\rm s}$CDM simultaneously ameliorates the $H_0$, $M_{\rm B}$, and $S_8$ tensions, though the inclusion of full BAO data in the analysis compromises the model's success by moving $z_{\dagger}$ to $\sim2.4$---in this case the negative cosmological constant, $\Lambda_{\rm s}(z>z_\dagger)=-\Lambda_{\rm s0}$, does not have enough time to significantly influence the evolution of the universe against the dust still dominating the universe.

This analysis was elaborated by using the Pantheon SNIa data~\cite{Pan-STARRS1:2017jku} (to break the degeneracy between $H_0$ and $z_\dagger$, without using BAO), both with and without the SH0ES $M_{\rm B}$ prior~\cite{Camarena:2021jlr}, and/or the \textit{completed} BAO data~\cite{eBOSS:2020yzd} along with Planck CMB in~\cite{Akarsu:2022typ}. It was shown that when the $M_{\rm B}$ prior is utilized without the full BAO data, predicting  $z_\dagger\sim1.8$, $\Lambda_{\rm s}$CDM reduces all the major discrepancies (related to $H_0$, $M_{\rm B}$, and $S_8$) that prevail within $\Lambda$CDM (and its canonical extensions) below $\sim1\sigma$, and is very strongly favored over $\Lambda$CDM in terms of Bayesian evidence. It is worth noting that the presence of sign-switching at $z_\dagger\sim2$ was originally motivated by BAO Ly-$\alpha$ (which favors negative/zero DE densities for $z\gtrsim2$) in~\cite{Akarsu:2019hmw} and then this was further supported in~\cite{Akarsu:2021fol} by showing that while the BAO Ly-$\alpha$ insists on $z_\dagger\lesssim2.3$, the galaxy BAO from $z_{\rm eff}=0.38$ is in opposition to this pushing $z_\dagger$ to values larger than 2. Pleasantly, the results in~\cite{Akarsu:2022typ} show that the presence of the $M_{\rm B}$ prior finds excellent constraints of $z_\dagger\sim2$ ($z_\dagger\sim1.8$ when full BAO data is not included) even when BAO Ly-$\alpha$ is not used. A close inspection of the fact that using full BAO data hinders the success of $\Lambda_{\rm s}$CDM in~\cite{Akarsu:2022typ} shows that $\Lambda_{\rm s}$CDM is discrepant with the galaxy BAO data from $z_{\rm eff}=0.15$ and $0.38$. Accordingly, to reconcile $\Lambda_{\rm s}$CDM with these  two galaxy BAO data as well, adding correction to it might be an option (e.g., by introducing wavelet-type corrections to the Hubble radius of $\Lambda_{\rm s}$CDM at low redshifts, as suggested in~\cite{Akarsu:2022lhx}), which, however, would come at the cost of introducing additional free parameters. Alternatively, this can be related to the discordance between low- and high-redshift BAO measurements in $\Lambda_{\rm s}$CDM, which is also present in $\Lambda$CDM~\cite{Aubourg:2014yra,duMasdesBourboux:2020pck}.

In this paper, we further investigate the $\Lambda_{\rm s}$CDM model  by using updated and extended data. To achieve this, along with Planck CMB data~\cite{Planck:2019nip},
we utilize the recent Pantheon+ SNe Ia sample~\cite{Brout:2022vxf}, and additionally the 2D BAO data~\cite{Nunes:2020hzy,deCarvalho:2021azj}, as a less model dependent alternative to 3D BAO data used in previous analyses of  $\Lambda_{\rm s}$CDM. Furthermore, for a robust assessment of the resolution of the $S_8$ tension within $\Lambda_{\rm s}$CDM, we use cosmic shear measurements obtained from the latest public data release of the Kilo-Degree Survey (KiDS-1000)~\cite{KiDS:2020suj}. This inclusion allows us to robustly determine its consistency with regards to amplitude and growth of structures. 
By combining these diverse data sets, we aim to provide a comprehensive analysis that further establishes the viability and robustness of $\Lambda_{\rm s}$CDM in addressing various cosmological tensions and discrepancies.

\textit{Data sets and Methodology} -- We describe below the observational data sets and the statistical methods used to explore the parameter space. \textit{CMB}: The full Planck 2018 temperature and polarization likelihood~\cite{Planck:2019nip} in combination with the Planck 2018 lensing likelihood~\cite{Planck:2018lbu}. We refer to this data set as Planck. 
\textit{Transversal BAO}: Measurements of 2D BAO, $\theta_{\text{BAO}}(z)$, obtained in a weakly model-dependent approach, compiled in Table I in~\cite{Nunes:2020hzy,deCarvalho:2021azj}. These measurements were obtained using public data releases (DR) of the Sloan Digital Sky Survey (SDSS), namely: DR7, DR10, DR11, DR12, DR12Q (quasars), and following the same methodology in all measurements. The main differences (methodology and sample) between the 3D and 2D BAO measurements are discussed in~\cite{Bernui_2023,Nunes_2020} and references therein. We refer to this data set as BAOtr. \textit{Type Ia supernovae and Cepheids}: We use the SNe Ia distance moduli measurements from the Pantheon+ sample~\cite{Brout:2022vxf}, which consists of 1701 light curves of 1550 distinct SNe Ia ranging in the redshift interval $z \in [0.001, 2.26]$. We refer to this dataset as PantheonPlus. We also consider the SH0ES Cepheid host distance anchors, which facilitate constraints on both $M_{\rm B}$ and $H_0$. When utilizing SH0ES Cepheid host distances, the SNe Ia distance residuals are modified following the relationship Eq.(14) of~\cite{Brout:2022vxf}. We refer to this dataset as PantheonPlus\&SH0ES. \textit{Cosmic Shear}: We use KiDS-1000 data~\cite{Kuijken:2019gsa,Giblin:2020quj}. This includes the weak lensing two-point statistics data for both the auto and cross-correlations across five tomographic redshift bins~\cite{Hildebrandt:2020rno}. We employ the public likelihood in \footnote{\href{https://github.com/BStoelzner/KiDS-1000_MontePython_likelihood}{KiDS-1000 Montepython likelihood}}. We follow the KiDS team analysis and adopt the COSEBIs (Complete Orthogonal Sets of E/B-Integrals) likelihood in our results~\cite{KiDS:2020suj}. For the prediction of the matter power spectrum, we use the augmented halo model code, HMcode~\cite{Mead:2015yca}. We highlight that at level of the linear perturbations theory and Boltzmann equations, $\Lambda_{\rm s}$CDM have exactly the same shape as predicted by $\Lambda$CDM. The only effect on the matter power spectrum comes from the $H(z)$ behavior at late times. As HMcode is robustly tested at percent level for variation on $H(z)$ functions beyond $\Lambda$CDM, we conclude that no further change on the HMcode is necessary to apply cosmic shear measurements on $\Lambda_{\rm s}$CDM. We refer to this data set as KiDS-1000. 

We explore the full parameter space of the $\Lambda_{\rm s}$CDM model and, for comparison, that of $\Lambda$CDM. The baseline seven free parameters of $\Lambda_{\rm s}$CDM are given by $\mathcal{P}= \left\{ \omega_{\rm b}, \, \omega_{\rm c}, \, \theta_s, \,  A_{\rm s}, \, n_s, \, \tau_{\rm reio},
\,   z_\dagger \right\}$, where the first six are the common ones with $\Lambda$CDM. We use \texttt{CLASS$+$MontePython} code~\cite{Blas:2011rf, Lesgourgues:2011re, Brinckmann:2018cvx} with Metropolis-Hastings mode to derive constraints on cosmological parameters for $\Lambda_{\rm s}$CDM baseline from several combinations of the data sets defined above, ensuring a Gelman-Rubin convergence criterion~\cite{Gelman:1992zz} of ${R-1 < 10^{-2}}$ in all the runs. For the model comparison, we compute the relative log-Bayesian evidence $\ln \mathcal{B}_{ij}$ to estimate the Evidence of $\Lambda_{\rm s}$CDM with respect to $\Lambda$CDM, through the publicly available package \texttt{MCEvidence} \footnote{\href{https://github.com/yabebalFantaye/MCEvidence}{github.com/yabebalFantaye/MCEvidence}}~\cite{Heavens:2017hkr,Heavens:2017afc}. We use the convention of a negative value if $\Lambda_{\rm s}$CDM is preferred against $\Lambda$CDM, or vice versa, and we refer to the revised Jeffreys' scale by Trotta~\cite{Kass:1995loi,Trotta:2008qt}, to interpret the results. We will say that the evidence is inconclusive if $0 \leq | \ln \mathcal{B}_{ij}|  < 1$, weak if $1 \leq | \ln \mathcal{B}_{ij}|  < 2.5$, moderate if $2.5 \leq | \ln \mathcal{B}_{ij}|  < 5$, strong if $5 \leq | \ln \mathcal{B}_{ij}|  < 10$, and very strong if $| \ln \mathcal{B}_{ij} | \geq 10$.

\begin{table*}[ht!]
     \caption{Marginalized constraints, mean values with 68\% CL (bestfit value), on  the free and some derived parameters of the $\Lambda_{\rm s}$CDM and standard $\Lambda$CDM models for different data set combinations. Bayes factors $\mathcal{B}_{ij}$ given by $\ln \mathcal{B}_{ij} = \ln \mathcal{Z}_{\Lambda \rm CDM} - \ln \mathcal{Z}_{\Lambda_{\rm s} \rm CDM}$ are also displayed for the different analyses so that a  negative value indicates a preference for the $\Lambda_{\rm s}$CDM model against the $\Lambda$CDM scenario.  }
     \label{tab:results}
     \scalebox{0.7}{
 \begin{tabular}{lccccc}
  	\hline
    \toprule
    \textbf{Data set } & \textbf{Planck}& \textbf{Planck+BAOtr} & \textbf{Planck+BAOtr} \;\; & \textbf{Planck+BAOtr}\;\; & \textbf{Planck+BAOtr}   \\
 &  & & \textbf{+PP} \;\; & \textbf{+PP\&SH0ES}\;\; & \textbf{+PP\&SH0ES+KiDS-1000}   \\  \hline
      \textbf{Model} & \textbf{$\bm{\Lambda}_{\textbf{s}}$CDM}&
       \textbf{$\bm{\Lambda}_{\textbf{s}}$CDM}&\textbf{$\bm{\Lambda}_{\textbf{s}}$CDM}&\textbf{$\bm{\Lambda}_{\textbf{s}}$CDM}&\textbf{$\bm{\Lambda}_{\textbf{s}}$CDM}\\
        & \textcolor{blue}{\textbf{$\bm{\Lambda}$CDM}} & \textcolor{blue}{\textbf{$\bm{\Lambda}$CDM}} & \textcolor{blue}{\textbf{$\bm{\Lambda}$CDM}} & \textcolor{blue}{\textbf{$\bm{\Lambda}$CDM}} & \textcolor{blue}{\textbf{$\bm{\Lambda}$CDM}} 
          \\ \hline
      \vspace{0.1cm}

{\boldmath$z_{\dagger}             $}& unconstrained & $1.70^{+0.09}_{-0.19}(1.65)$& $1.87^{+0.13}_{-0.21}(1.75)$ & $1.70^{+0.10}_{-0.13}(1.67) $ & $1.72^{+0.09}_{-0.12}(1.70)$\\

& \textcolor{blue}{$--$} &\textcolor{blue}{$--$}      & \textcolor{blue}{$--$}  & \textcolor{blue}{$--$} & \textcolor{blue}{$--$}\\

\hline

{\boldmath$M_B    {\rm[mag]}        $}& $--$&$ --$& $-19.317^{+0.021}_{-0.025}(-19.311)$& $-19.290\pm 0.017(-19.278)$ & $-19.282\pm 0.017(-19.280) $\\
 & \textcolor{blue}{$--$ } & \textcolor{blue}{$--$ }      & \textcolor{blue}{$-19.407\pm 0.013(-19.411)$} & \textcolor{blue}{$-19.379\pm 0.012(-19.373)$} & \textcolor{blue}{$-19.372\pm 0.011(-19.369)$}\vspace{0.2cm} \\

{\boldmath$H_0 {\rm[km/s/Mpc]}            $}& $70.77^{+0.79}_{-2.70}  ( 71.22)$& $73.30^{+1.20}_{-1.00}(73.59)$& $71.72^{+0.73}_{-0.92}(71.97) $ & $72.82\pm 0.65(73.20) $& $73.16\pm 0.64(73.36)$\\
 & \textcolor{blue}{$67.39\pm 0.55 (67.28)$ }&\textcolor{blue}{$68.84\pm 0.48(68.61)$}       & \textcolor{blue}{$68.55\pm 0.44( 68.54)$} & \textcolor{blue}{$69.57\pm 0.42 (69.73) $} & \textcolor{blue}{$69.83\pm 0.37(69.96)$}\vspace{0.2cm} \\


{\boldmath$\Omega_{\rm m}  $}& $0.2860^{+0.0230}_{-0.0099}(0.2796)$& $0.2643^{+0.0072}_{-0.0090}(0.2618)$& $0.2768^{+0.0072}_{-0.0063}(0.2759)$ & $0.2683\pm 0.0052( 0.2646)$ & $0.2646\pm 0.0052(0.2622)$\\ 
 & \textcolor{blue}{$0.3151\pm 0.0075 (0.3163)$}  &\textcolor{blue}{$0.2958\pm 0.0061(0.2984)$}      & \textcolor{blue}{$0.2995\pm 0.0056(0.2992)$} & \textcolor{blue}{$0.2869\pm 0.0051(0.2849) $}& \textcolor{blue}{$0.2837\pm 0.0045 (0.2816) $}  \vspace{0.2cm}  \\


{\boldmath$S_8             $}& $0.801^{+0.026}_{-0.016}(0.791) $& $0.777\pm 0.011(0.772)$& $0.791\pm 0.011(0.794)$ & $0.783\pm 0.010(0.777)$& $0.774\pm 0.009(0.773)$\\
& \textcolor{blue}{$0.832\pm 0.013(0.835)$}  &\textcolor{blue}{$0.802\pm 0.011 (0.804)$}      & \textcolor{blue}{$0.808\pm 0.010(0.804)$} & \textcolor{blue}{$0.788\pm 0.010 (0.784)$}  & \textcolor{blue}{$0.781\pm 0.008(0.782) $} \\



\hline


 {\boldmath$\chi^2_{\rm min}$}& $2778.06$& $2793.38$& $4219.68$& $4097.32$& $4185.34$\\
 & \textcolor{blue}{$2780.52$}  & \textcolor{blue}{$2820.30$}  & \textcolor{blue}{$4235.18$} & \textcolor{blue}{$4138.26$}  & \textcolor{blue}{$4226.50$} \\


{\boldmath${\rm ln} \mathcal{B}_{ij}$}& 
 $-1.28$   &$-12.65$  & $-7.52$ & $-19.47$ & $-19.77$ \\
 \hline
 \hline
\end{tabular}
}
\end{table*}

\begin{figure}[t!]
    \centering
    \includegraphics[width=7.6cm]{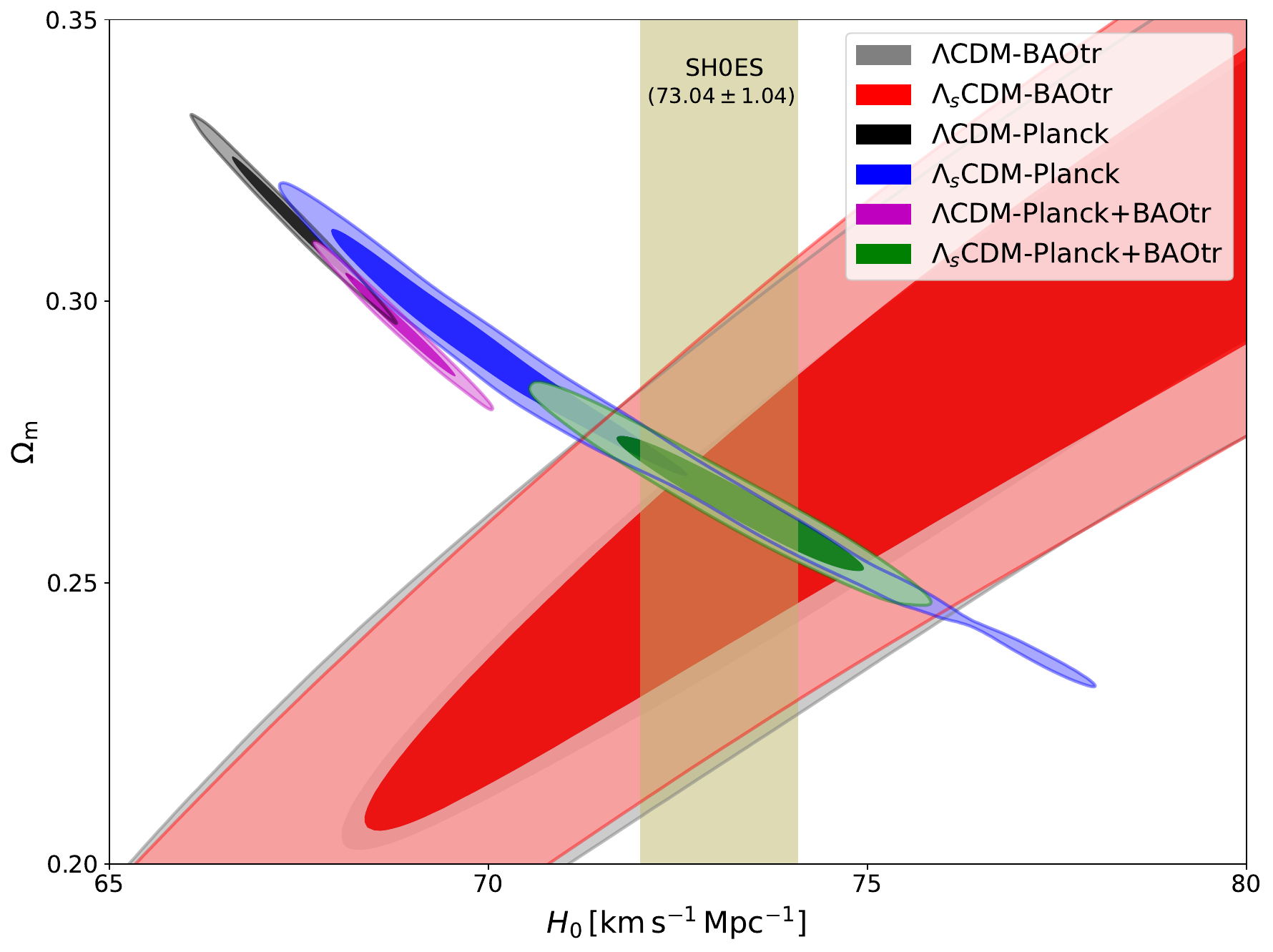}
    \caption{2D contours at 68\% and 95\% CLs in the $H_0$-$\Omega_{\rm m}$ plane for the $\Lambda_{\rm s}$CDM and $\Lambda$CDM models from the Planck and/or BAOtr data.  It deserves mention that, in case of $\Lambda_{\rm s}$CDM, the Planck and BAOtr contours intersect right on the vertical band of SH0ES measurement.}
    \label{fig:BAOtr}
\end{figure}

\emph{Results} -- We present, in~\cref{tab:results}, the 68\% CL constraints on the main cosmological parameters of interest of the $\Lambda_{\rm s}$CDM and $\Lambda$CDM models obtained in our analyses by using different combinations of data sets, while we provide the complete table for the entire parameter space of the two models in the \textit{Supplemental Material}. When we consider only Planck data, we notice that the characteristic parameter of the $\Lambda_{\rm s}$CDM model,  $z_\dagger$, remains unconstrained, and we find strong degeneracy with other derived parameters, especially with $H_0$ and $\Omega_{\rm m}$, increasing values in $H_0$ and decreasing the total matter density parameter today. To break the degeneracy, we include the BAOtr data in our analysis with Planck data, which enables the constraint: $z_\dagger = 1.70^{+0.09}_{-0.19}$. Interestingly, this inclusion of BAOtr data leads to a higher value of the Hubble constant, viz., $H_0 = 73.3^{+1.2}_{-1.0}{\rm \,km\, s^{-1}\, Mpc^{-1}}$ which is perfectly consistent with the SH0ES measurement $H_0=73.04\pm1.04~{\rm km\, s^{-1}\, Mpc^{-1}}$~\cite{Riess:2021jrx}. In particular, it is noteworthy to observe in~\cref{fig:BAOtr} that the two models yield almost the identical contours for the BAOtr data (along with BBN prior $10^2\omega_{\rm b}^{\rm LUNA}=2.233\pm0.036$~\cite{Mossa:2020gjc}), while the BAOtr and Planck contours disagree in the case of $\Lambda$CDM; however, when considering $\Lambda_{\rm s}$CDM, it is striking that the BAOtr and Planck  contours precisely intersect at the vertical band of SH0ES $H_0$ measurement. Considering this remarkable success of $\Lambda_{\rm s}$CDM in addressing the $H_0$ tension, we proceed to incorporate the new PantheonPlus (PP) sample into our analysis, both with and without the Cepheids calibration provided by SH0ES. From the combination of the Planck, BAOtr, and PantheonPlus data sets, we find the constraints: $z_\dagger=1.87^{+0.13}_{-0.21}$ and $H_0 = 71.72^{+0.73}_{-0.92}{\rm \,km\, s^{-1}\, Mpc^{-1}}$. This constraint on $H_0$ is again consistent with the SH0ES measurement. Based on this finding, we confidently include the calibration provided by SH0ES, leaving our conclusions unchanged. We also note that the discrepancy in $M_{\rm B}$ between the SH0ES data ($M_B=-19.244\pm0.037\,{\rm mag}$) and the base $\Lambda$CDM cosmology inferred from Planck ($M_B=-19.401\pm0.027\,{\rm mag}$) is here resolved within the framework of $\Lambda_{\rm s}$CDM model. Thus, the $\Lambda_{\rm s}$CDM model also provides a robust solution to the $M_{\rm B}$ tension.

\begin{figure}[b!]
    \centering
    \includegraphics[width=4.24cm]{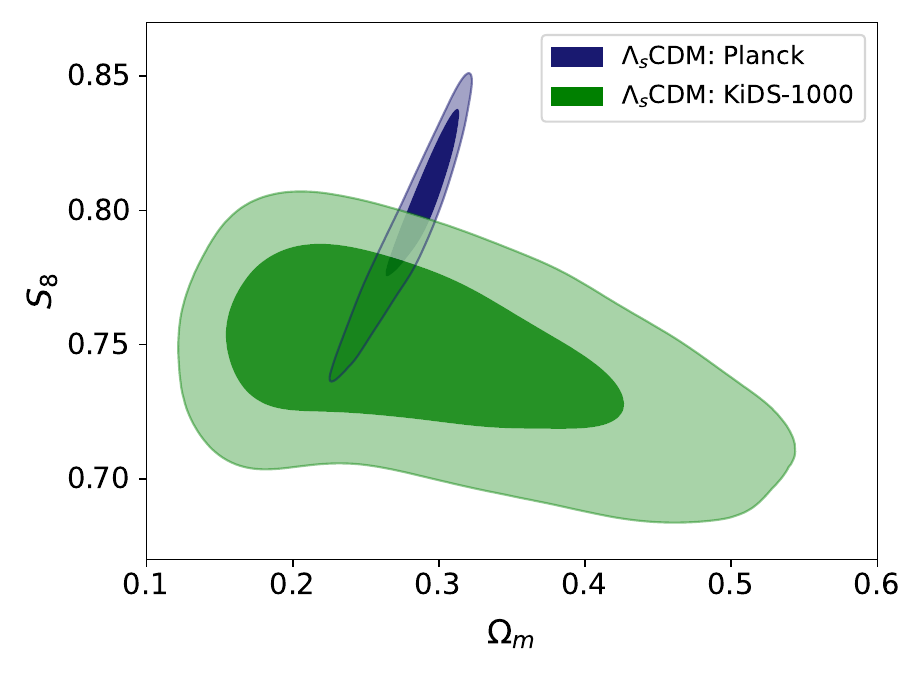} 
  \includegraphics[width=4.24cm]{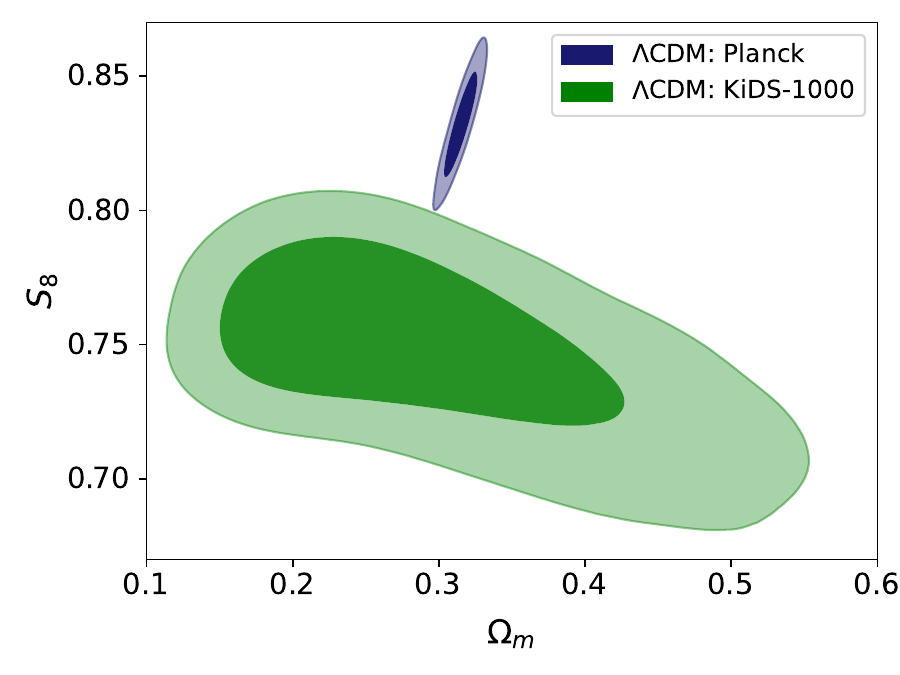}
    \includegraphics[width=4.24cm]{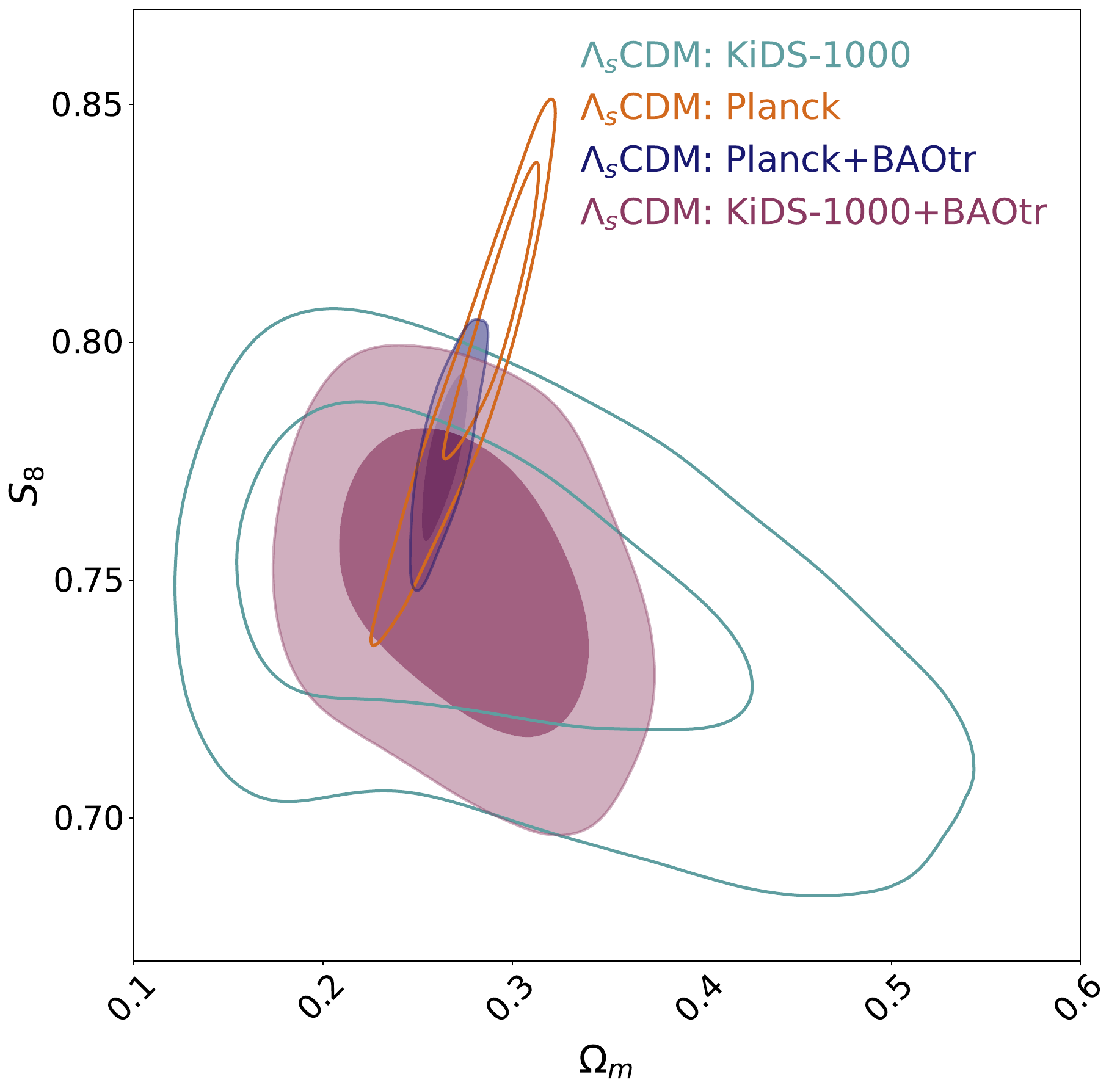} 
   \includegraphics[width=4.24cm]{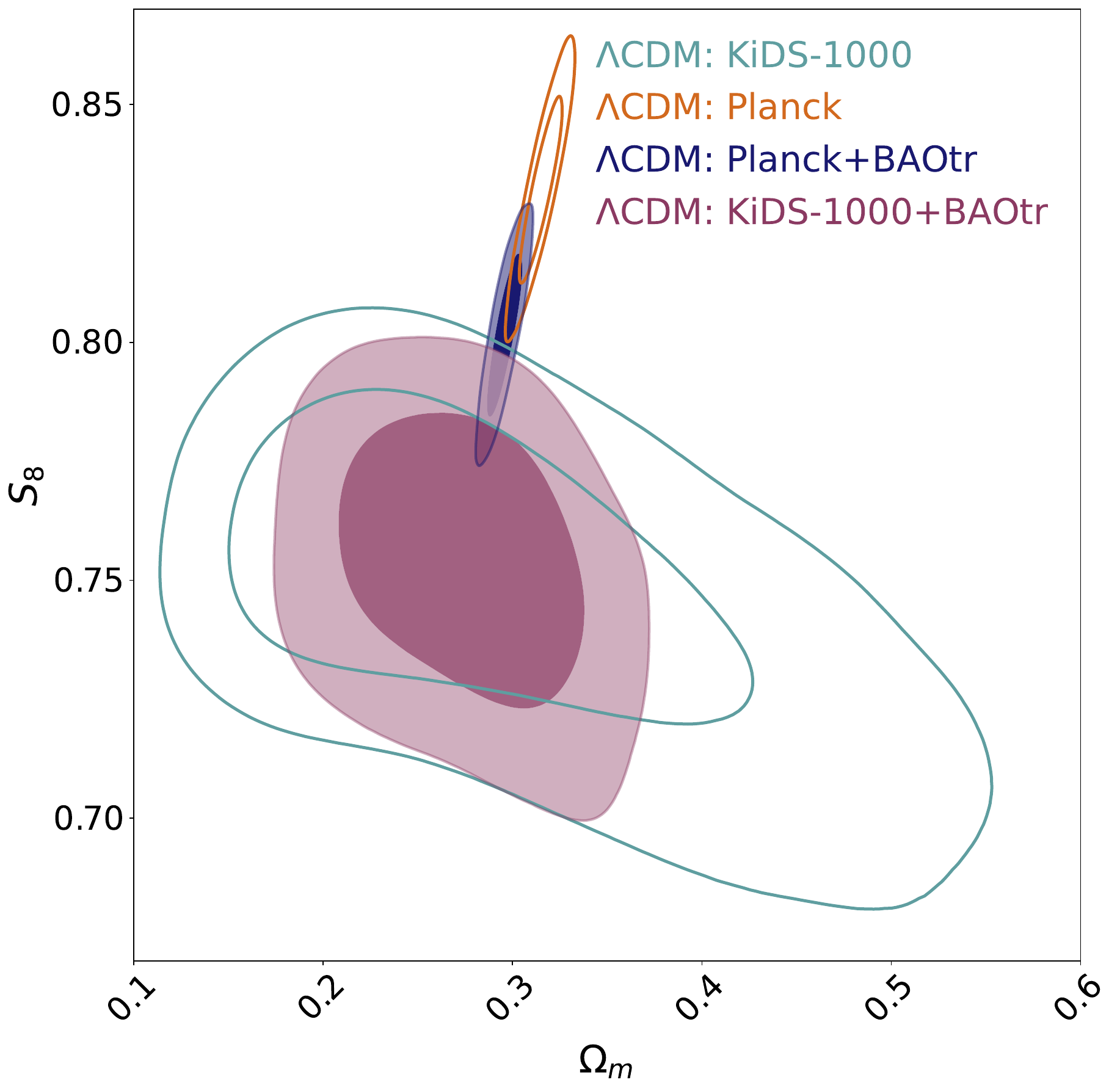}
    \caption{2D contours at 68\% and 95\% CLs in the $\Omega_{\rm m}$-$S_8$ plane for the $\Lambda_{\rm s}$CDM and $\Lambda$CDM models. $S_8 = 0.801^{+0.026}_{-0.016} $ ($\Lambda_{\rm s}$CDM: Planck),  $S_8 = 0.746^{+0.026}_{-0.021}  $ ($\Lambda_{\rm s}$CDM: KiDS-1000), $S_8 = 0.832\pm 0.013  $ ($\Lambda$CDM: Planck),   $S_8 =0.749^{+0.027}_{-0.020}$ ($\Lambda$CDM: KiDS-1000) at 68\% CL. }
    \label{fig:kids}
\end{figure}

From the results of these analyses, we further notice that $S_8$ and $\Omega_{\rm m}$ get lower values in $\Lambda_{\rm s}$CDM compared to $\Lambda$CDM.  In order to see whether $S_8$ tension is resolved in our model, we separately analyse the models with KiDS-1000 only data. The upper panels of~\cref{fig:kids} show 2D contours at 68\%, and 95\% CLs in the $\Omega_{\rm m}$-$S_8$ plane for the $\Lambda_{\rm s}$CDM and $\Lambda$CDM models. We note that the two models yield very similar contours for the KiDS data (as both models being the same at redshifts relevant to KiDS data). However, while the Planck and KiDS contours disagree in $\Lambda$CDM, they do agree in $\Lambda_{\rm s}$CDM as the Planck contour extends directly into the KiDS contour. This occurs because the smaller $z_\dagger$ values allowed by Planck data lead to smaller values of $S_8$ and $\Omega_{\rm m}$. For $\Lambda_{\rm s}$CDM, the observational constraints on $S_8$, viz., $S_8 = 0.801^{+0.026}_{-0.016} $ from Planck and $S_8 = 0.746^{+0.026}_{-0.021} $ from KiDS-1000, are compatible with each other, and thus the $S_8$ tension does not exist in $\Lambda_{\rm s}$CDM while it prevails in $\Lambda$CDM. The contour plots for $\Lambda_{\rm s}$CDM in the lower left panel of~\cref{fig:kids} (see the lower right panel for $\Lambda$CDM) further show the robustness of the constraints on $S_8$ in the presence of other data sets under consideration.

\begin{figure*}[t!]
    \centering
    \includegraphics[width=12.0cm]{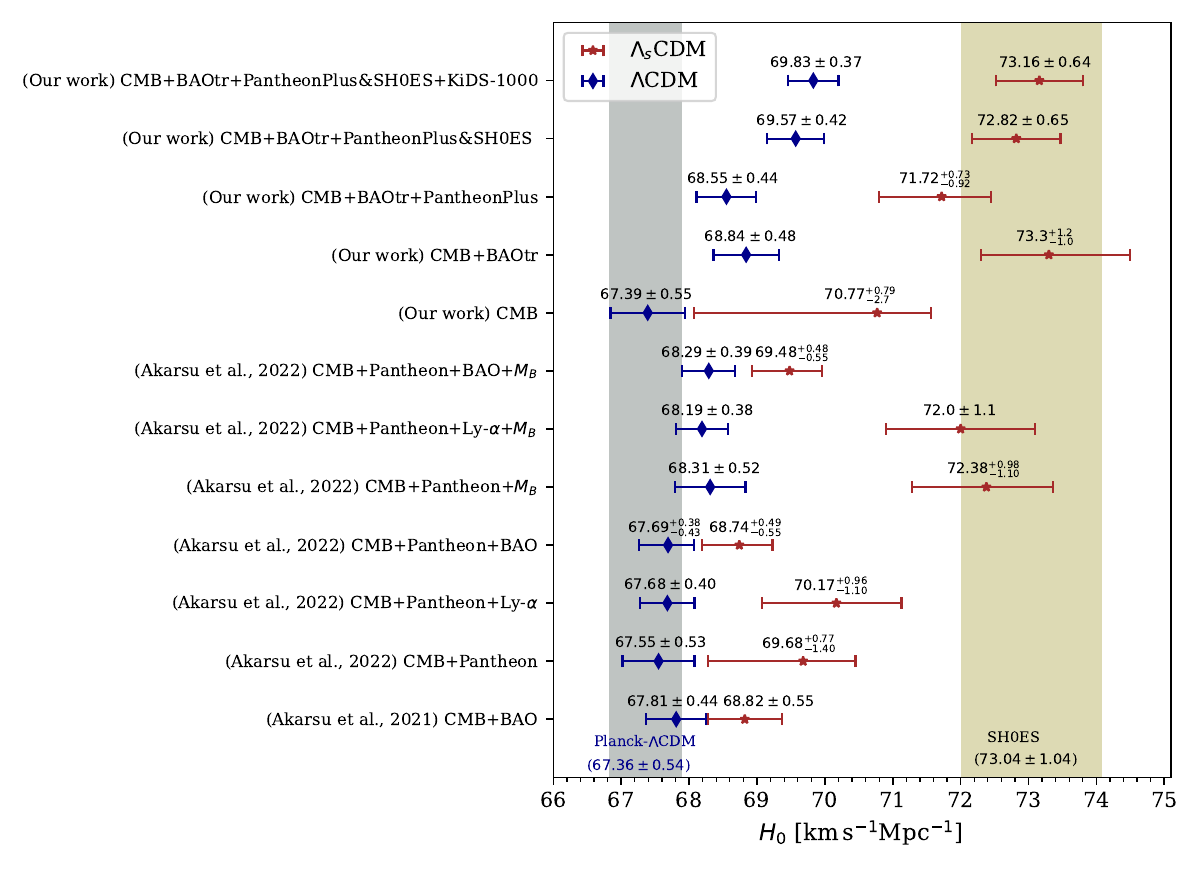} 
    \caption{The mean values with 68\% CL on $H_0$ for the $\Lambda_{\rm s}$CDM and $\Lambda$CDM models from various data combinations. The left vertical band stands for Planck-$\Lambda$CDM constraint: $H_0=67.36\pm0.54{\rm \,km\, s^{-1}\, Mpc^{-1}}$, and the right one is for the latest SH0ES measurement~\cite{Riess:2021jrx}: $H_0=73.04\pm1.04{\rm \,km\, s^{-1}\, Mpc^{-1}}$. We see that there is no $H_0$ tension in most data combinations for $\Lambda_{\rm s}$CDM, in particular, when we use the less model-dependent BAOtr data. The only exception is in the cases explored in previous works~\cite{Akarsu:2021fol,Akarsu:2022typ} that employ the 3D BAO data (BAO in the plot), among which the two galaxy BAO data from $z_{\rm eff}=0.15$ and $0.38$ preventing the model from successfully resolving the tension.}
    \label{plot:whisker}
\end{figure*}

Finally, from the analysis of the models using the combination of all the data sets under consideration, viz., Planck+BAOtr+PP\&SH0ES+KiDS-1000, we obtain the most robust constraints on the model parameters as shown in the last column of~\cref{tab:results}.~\cref{plot:whisker} shows the whisker plot displaying 68\% CL constraints on $H_0$ for the $\Lambda_{\rm s}$CDM and $\Lambda$CDM models from various data combinations. We see that there is no  $H_0$  tension in the present analyses of $\Lambda_{\rm s}$CDM with all data combinations including the BAOtr data. We emphasize that all data sets under consideration are compatible within the framework of $\Lambda_{\rm s}$CDM. Note that there is no $H_0$ tension in our results using the BAOtr data, while using 3D BAO data $\Lambda_{\rm s}$CDM reduces the $H_0$ tension but fails to fully resolve it in previous works~\cite{Akarsu:2021fol,Akarsu:2022typ}. We attribute this to the possible model dependence of the 3D BAO reconstruction, which is (mostly) absent for the BAOtr data. In particular, \cite{Akarsu:2022typ} finds that the two galaxy 3D BAO data---viz., SDSS main galaxy sample (MGS) from $z_{\rm eff}=0.15$ and BOSS galaxy from $z_{\rm eff}=0.38$---are responsible for this reduced concordance under $\Lambda_{\rm s}$CDM.

In addition, as can be seen in \textit{Supplemental Material}, $\Lambda_{\rm s}$CDM poses no problems with any of the well-known parameters of the Universe based on observations and, on theoretical side, standard physics. Instead, the constraints on its six baseline parameters that are common with $\Lambda$CDM, show unprecedented stability in the face of different data sets. Moreover, $\Lambda_{\rm s}$CDM is consistent with BAO Ly-$\alpha$ data. It predicts the age of the universe, $t_0=13.522\pm 0.027$ Gyr, which is consistent with estimations utilizing the oldest globular clusters, e.g., $t_{\rm u}=13.50\pm0.15({\rm stat.})$ Gyr~\cite{Valcin:2021jcg}. Furthermore, it maintains the physics and dynamics of the pre-recombination universe as they are in the standard model. For example, the constraints on the drag redshift and the sound horizon at this epoch ($r_{\rm d}$ and $z_{\rm d}$) remain unaltered compared to those obtained within $\Lambda$CDM. Additionally, the constraints on $Y_{\rm P}$ (the primordial mass fraction of $^4\rm He$) and $\omega_{\rm b}$ (the present-day physical density parameter of baryons) are consistent with the standard BBN.

To provide a conclusive assessment of the robustness of our results, we conduct a Bayesian model comparison to evaluate the relative performance of $\Lambda_{\rm s}$CDM and $\Lambda$CDM in terms of their statistical fit to the data. The results of the relative Bayesian evidence are presented in the lower section of~\cref{tab:results}. According to the revised Jeffreys' scale, the evidence in favor of the $\Lambda_{\rm s}$CDM scenario is found \textit{weak} when considering the Planck data alone. However, it strengthens to a \textit{strong} level in the analysis with Planck+BAOtr+PP data. Remarkably, for all the other combinations of data-sets considered here, the evidence in favor of the $\Lambda_{\rm s}$CDM scenario turns out be \textit{very strong}. Thus, $\Lambda_{\rm s}$CDM finds by far better statistical fit to the data compared to $\Lambda$CDM.

The findings in this study indicate that the $\Lambda_{\rm s}$CDM model consistently outperforms the standard $\Lambda$CDM model not only in resolving the prominent cosmological tensions, but also in terms of the statistical fit to the data across various data-set combinations, providing very strong support for its validity and effectiveness in explaining the observed cosmological phenomena. In addition, $\Lambda_{\rm s}$CDM mitigates several other tensions of lower statistical significance, as illustrated in \textit{Supplemental Material}, where we also provide additional information about the main results discussed here.

\emph{Final remarks} -- Using the state-of-the-art methodology for the observational constraints and recent data available in the literature, we show that a simple model, viz., $\Lambda_{\rm s}$CDM~\cite{Akarsu:2019hmw,Akarsu:2021fol,Akarsu:2022typ}, which experiences a rapid transition of the Universe from anti-de Sitter vacua to de Sitter vacua (namely, the cosmological constant switches sign from negative to positive) at late times ($z_\dagger\approx1.7$), can address the major cosmological tensions ($H_0$, $M_{\rm B}$, and $S_8$ tensions) simultaneously; in particular, when we use BAOtr data, which are less model-dependent, unlike the previous works on $\Lambda_{\rm s}$CDM that used 3D BAO data. Our proposal consists of the most economical cosmological model available in the literature with that ability, because it does not involve any extra physical parameters beyond $\Lambda$CDM, but only a cosmic time transition which needs to be fixed by data. The abrupt/rapid nature of the $\Lambda_{\rm s}$, or a dark energy mimicking it, and also the fact that it shifts from negative to positive value, may render finding a concrete physical mechanism underlying this model challenging. However, the phenomenological success of $\Lambda_{\rm s}$CDM despite its simplicity, is highly encouraging to look for possible underlying physical mechanisms; see~\cite{Akarsu:2019hmw,Akarsu:2021fol,Akarsu:2022typ} for further discussions and~\cite{Alexandre:2023nmh} realizing abrupt sign-switching cosmological constant via a classical metric signature change across boundaries with a degenerate metric in different formulations of general relativity. On the other hand, our findings may have far-reaching implications in theoretical physics as negative cosmological constant is a theoretical sweet spot--viz., AdS space/vacuum is welcome due to the AdS/CFT correspondence~\cite{Maldacena:1997re} and is preferred by string theory and string theory motivated supergravities~\cite{Bousso:2000xa}. And thereby, it would be natural to associate this phenomenon to a possible (phase) transition from AdS to dS that is derived in such fundamental theories of physics, and the theories that find motivation from them. Thus, it is essential to conduct further research on refinement at both theoretical and observational level, particularly on identifying observables unique to this model and searching for their traces in the sky, for establishing the $\Lambda_{\rm s}$CDM model as a promising candidate or guide for a new concordance cosmological model of the Universe.

\emph{Acknowledgements} -- This work is dedicated to the memory of Professor John David Barrow. We gratefully acknowledge Marika Asgari for the valuable discussions, inputs in the paper, and help with the KiDS data analysis. We thank Joseph Silk for useful discussions. \"{O}.A. acknowledges the support by the Turkish Academy of Sciences in scheme of the Outstanding Young Scientist Award  (T\"{U}BA-GEB\.{I}P). \"{O}.A. is supported in part by TUBITAK grant 122F124. E.D.V. is supported by a Royal Society Dorothy Hodgkin Research Fellowship. S.K. gratefully acknowledges support from the Science and Engineering Research Board (SERB), Govt. of India (File No.~CRG/2021/004658). 
R.C.N. thanks the CNPq for partial financial support under the project No. 304306/2022-3. J.A.V. acknowledges the support provided by FOSEC SEP-CONACYT Investigaci\'on B\'asica A1-S-21925, Ciencias de Frontera
CONACYT-PRONACES/304001/202 and UNAM-DGAPA-PAPIIT IN117723. A.Y. is supported by Junior Research Fellowship (CSIR/UGC Ref. No. 201610145543) from University Grants Commission, Govt. of India. This article is based upon work from COST Action CA21136 Addressing observational tensions in cosmology with systematics and fundamental physics (CosmoVerse) supported by COST (European Cooperation in Science and Technology).

\bibliography{lscdm}
\end{document}


\title{Supplemental material}

\author{\"{O}zg\"{u}r Akarsu}
\email{akarsuo@itu.edu.tr}
\affiliation{Department of Physics, Istanbul Technical University, Maslak 34469 Istanbul, Turkey}

\author{Eleonora Di Valentino}
\email{e.divalentino@sheffield.ac.uk}
\affiliation{School of Mathematics and Statistics, University of Sheffield, Hounsfield Road, Sheffield S3 7RH, United Kingdom}

\author{Suresh Kumar}
\email{suresh.math@igu.ac.in}
\affiliation{Department of Mathematics, Indira Gandhi University, Meerpur, Haryana 122502, India}

\author{Rafael C. Nunes}
\email{rafadcnunes@gmail.com}
\affiliation{Instituto de F\'{i}sica, Universidade Federal do Rio Grande do Sul, 91501-970 Porto Alegre RS, Brazil}
\affiliation{Divis\~ao de Astrof\'isica, Instituto Nacional de Pesquisas Espaciais, Avenida dos Astronautas 1758, S\~ao Jos\'e dos Campos, 12227-010, SP, Brazil}

\author{J. Alberto Vazquez}
\email{javazquez@icf.unam.mx}
\affiliation{Instituto de Ciencias F\'isicas, Universidad Nacional Aut\'onoma de M\'exico, Cuernavaca, Morelos, 62210, M\'exico}

\author{Anita Yadav}
\email{anita.math.rs@igu.ac.in }
\affiliation{Department of Mathematics, Indira Gandhi University, Meerpur, Haryana 122502, India}

\keywords{}

\pacs{}

\maketitle

\section{Model Baseline and prior info}

\begin{figure*}
    \centering
    \includegraphics[width=17cm]{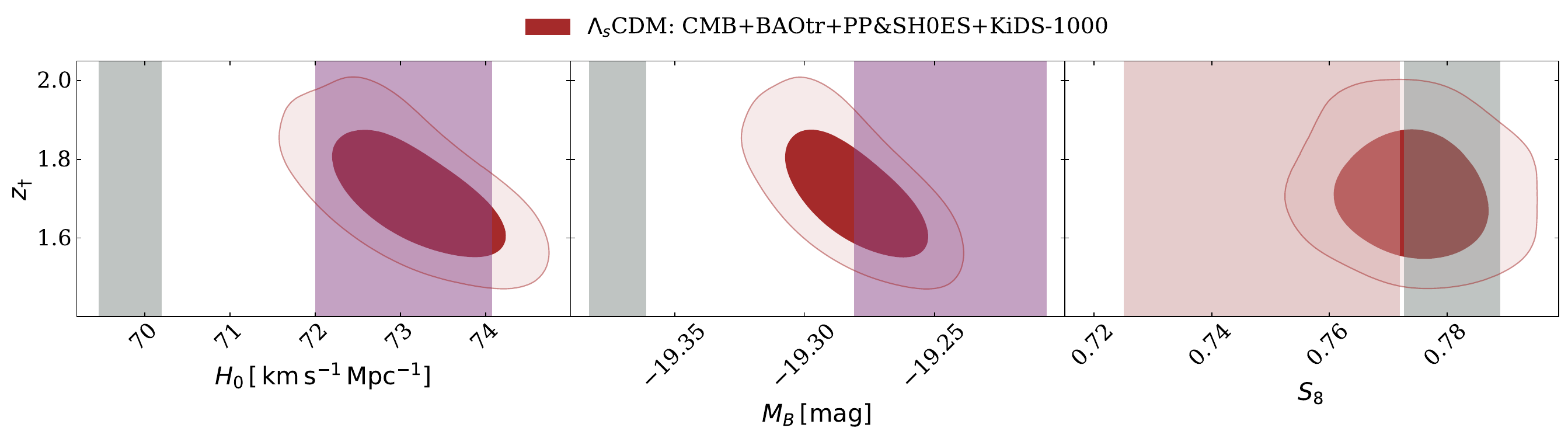}
    \includegraphics[width=17cm]{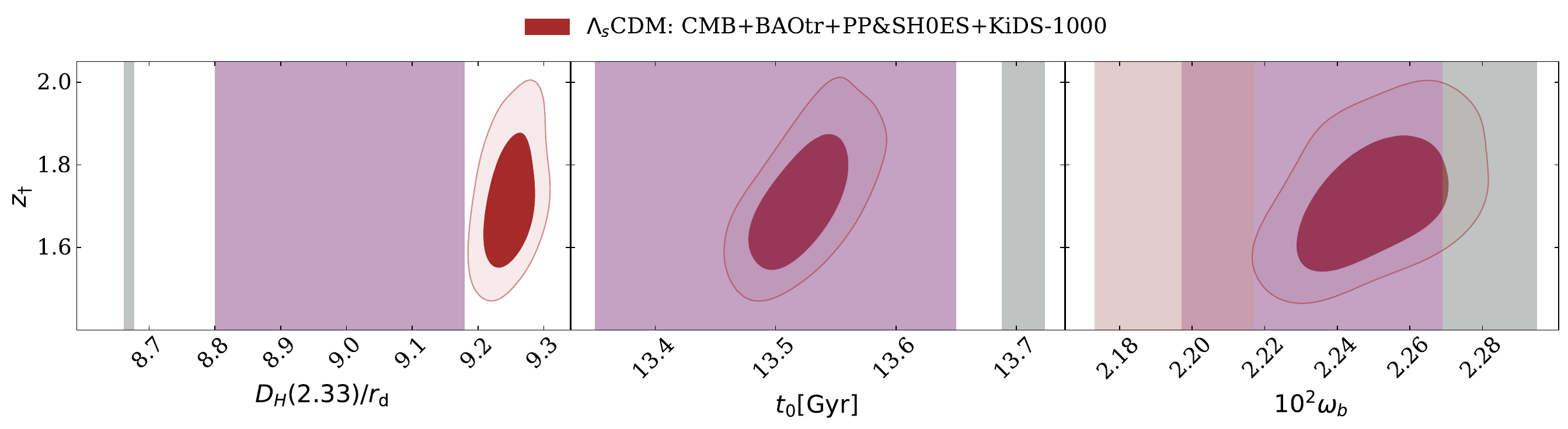} 
 \caption{Two-dimensional marginalized probability posteriors of $z_{\dagger}$ versus $H_0$, $M_{\rm B}$, $S_8$, $D_H(2.33)/r_{\rm d}$ ($D_H/r_{\rm d}$ at $z_{\rm eff}=2.33$ relevant to the Ly-$\alpha$ measurements), $t_0$, and $\omega_{\rm b}$ in $\Lambda_{\rm s}$CDM model for CMB+BAOtr+PP$\&$SH0ES+KiDS-1000. The vertical grey bands are the constraints (68\% CL) for the $\Lambda$CDM model. The vertical purple bands stand for the theoretical/direct observational estimations (at 68\% CL) of the corresponding parameters commonly used in the literature: $H_{\rm 0}=73.04\pm1.04~{\rm km\, s^{-1}\, Mpc^{-1}}$(SH0ES)~\cite{Riess:2021jrx}; $M_{\rm B}=-19.244\pm 0.037\,\rm mag$ (SH0ES)~\cite{Camarena:2021dcvm};  $D_H(2.33)/r_{\rm d}=8.99\pm0.19$ (for combined Ly-$\alpha$ data)~\cite{duMasdesBourboux:2020pck}; $t_{0}=13.50\pm0.15\,\rm Gyr$ (systematic uncertainties are not included)~\cite{Valcin:2021jcg}; $10^2\omega_{\rm b}^{\rm LUNA}=2.233\pm0.036$~\cite{Mossa:2020gjc}. In addition, we show vertical  brown band for $10^2\omega_{\rm b}^{\rm PCUV21}=2.195\pm0.022$~\cite{Pitrou:2020etk} and $S_8 = 0.746_{-0.021}^{+0.026}$ ($\Lambda_{\rm s}$CDM: KiDS-1000) [this work].}
\label{fig:zdagger2d} 
\end{figure*}

The baseline seven free parameters of $\Lambda_{\rm s}$CDM are given by $\mathcal{P}= \left\{ \omega_{\rm b}, \, \omega_{\rm c}, \, \theta_s, \,  A_{\rm s}, \, n_s, \, \tau_{\rm reio},
\,   z_\dagger \right\}$, where the first six are the common ones with $\Lambda$CDM, viz., $\omega_{\rm b}=\Omega_{\rm b} h^2$ and $\omega_{\rm c}=\Omega_{\rm c}h^2$ ($\Omega$ being the present-day density parameter) are, respectively, the present-day physical density parameters of baryons and CDM, $\theta_{\rm s}$ is the ratio of the sound horizon to the angular diameter distance at decoupling, $A_{\rm s}$ is the initial super-horizon amplitude of curvature perturbations at $k=0.05$ Mpc$^{-1}$, $n_{\rm s}$ is the primordial spectral index, and $\tau_{\rm reio}$ is the reionization optical depth. We assume three neutrino species, approximated as two massless states and a single massive neutrino of mass $m_{\nu}=0.06\,\rm eV$. We use uniform priors $\omega_{\rm b}\in[0.018,0.024]$, $\omega_{\rm c}\in[0.10,0.14]$, $100\,\theta_{\rm s}\in[1.03,1.05]$, $\ln(10^{10}A_{\rm s})\in[3.0,3.18]$, $n_{\rm s}\in[0.9,1.1]$, and $\tau_{\rm reio}\in[0.04,0.125]$ for the common free parameters, and $z_\dagger\in[1,3]$ for the additional free parameter of $\Lambda_{\rm s}$CDM. Also, we use uniform priors $S_8\in[0.1,1.3]$ and $h\in[0.64,0.82]$ for relevant analyses. In Table~\ref{tab:results}, we show observational constraints on the parameters of both the $\Lambda_{\rm s}$CDM and $\Lambda$CDM models from different data combinations. Fig.~\ref{fig:zdagger2d} displays two-dimensional marginalized probability posteriors of $z_{\dagger}$ versus $H_0$, $M_{\rm B}$, $S_8$, $D_H(2.33)/r_{\rm d}$ ($D_H/r_{\rm d}$ at $z_{\rm eff}=2.33$ relevant to the Ly-$\alpha$ measurements), $t_0$, and $\omega_{\rm b}$ in $\Lambda_{\rm s}$CDM model for CMB+BAOtr+PP$\&$SH0ES+KiDS-1000. Additionally, Table~\ref{tab:tensions} quantifies the concordance/discordance between the $\Lambda$CDM/$\Lambda_{\rm s}$CDM models and the theoretical/direct observational estimations, viz., $H_{\rm 0}=73.04\pm1.04~{\rm km\, s^{-1}\, Mpc^{-1}}$(SH0ES)~\cite{Riess:2021jrx}; $M_{\rm B}=-19.244\pm 0.037\,\rm mag$ (SH0ES)~\cite{Camarena:2021dcvm}; $D_H(2.33)/r_{\rm d}=8.99\pm0.19$ (for the combined Ly-$\alpha$ data)~\cite{duMasdesBourboux:2020pck}; $t_{0}=13.50\pm0.15\,\rm Gyr$ (systematic uncertainties are not included)~\cite{Valcin:2021jcg}; $10^2\omega_{\rm b}^{\rm LUNA}=2.233\pm0.036$ (empirical approach, based primarily on experimentally measured cross sections for $d(p,\gamma)^3\rm He$ reaction)~\cite{Mossa:2020gjc} and $10^2\omega_{\rm b}^{\rm PCUV21}=2.195\pm0.022$ (theoretical approach, incorporating nuclear theory for $d(p,\gamma)^3\rm He$ reaction)~\cite{Pitrou:2020etk}. $S_8 = 0.746^{+0.026}_{-0.021}  $ ($\Lambda_{\rm s}$CDM: KiDS-1000) and $S_8 =0.749^{+0.027}_{-0.020}$ ($\Lambda$CDM: KiDS-1000) obtained in this work.

\begin{table*}[t!]
     \caption{Marginalized constraints, mean values with 68\% CL (bestfit value), on the free and some derived parameters of the $\Lambda_{\rm s}$CDM and standard $\Lambda$CDM models for different data set combinations. Bayes factors $\mathcal{B}_{ij}$ given by $\ln \mathcal{B}_{ij} = \ln \mathcal{Z}_{\Lambda \rm CDM} - \ln \mathcal{Z}_{\Lambda_{\rm s} \rm CDM}$ are also displayed for the different analyses, so that a  negative value indicates a preference for the $\Lambda_{\rm s}$CDM model against the $\Lambda$CDM scenario.}
     \label{tab:results}
     \scalebox{0.72}{
 \begin{tabular}{lccccc}
  	\hline
    \toprule
    \textbf{Data set } & \textbf{Planck}& \textbf{Planck+BAOtr} & \textbf{Planck+BAOtr} \;\; & \textbf{Planck+BAOtr}\;\; & \textbf{Planck+BAOtr}   \\
 &  & & \textbf{+PP} \;\; & \textbf{+PP\&SH0ES}\;\; & \textbf{+PP\&SH0ES+KiDS-1000}   \\  \hline
      \textbf{Model} & \textbf{$\bm{\Lambda}_{\textbf{s}}$CDM}&
       \textbf{$\bm{\Lambda}_{\textbf{s}}$CDM}&\textbf{$\bm{\Lambda}_{\textbf{s}}$CDM}&\textbf{$\bm{\Lambda}_{\textbf{s}}$CDM}&\textbf{$\bm{\Lambda}_{\textbf{s}}$CDM}\\
        & \textcolor{blue}{\textbf{$\bm{\Lambda}$CDM}} & \textcolor{blue}{\textbf{$\bm{\Lambda}$CDM}} & \textcolor{blue}{\textbf{$\bm{\Lambda}$CDM}} & \textcolor{blue}{\textbf{$\bm{\Lambda}$CDM}} & \textcolor{blue}{\textbf{$\bm{\Lambda}$CDM}} 
          \\ \hline
      \vspace{0.1cm}

{\boldmath$10^{2}\omega{}_{b }$} & $2.241\pm 0.015(2.252) $&  $2.249\pm 0.014(2.251)$&$2.245\pm 0.014(2.247)$ & $2.246\pm 0.014(2.249)$ & $2.250\pm 0.013(2.252) $\\
 & \textcolor{blue}{$2.238\pm 0.014(2.235) $}  &  \textcolor{blue}{$2.262\pm 0.014(2.255)$}     & \textcolor{blue}{$2.256\pm 0.013(2.248)$} & \textcolor{blue}{$2.277\pm 0.013(2.280)$}  & \textcolor{blue}{$2.282\pm 0.013 (2.283)$} \\

{\boldmath$\omega{}_{cdm }$}& $0.1195\pm 0.0012(0.1187)$ &$0.1187\pm 0.0012 ( 0.1186)$ & $0.1192\pm 0.0011(0.1198)$ & $0.1192^{+0.0010}_{-0.0012}(0.1186)$& $0.1184\pm 0.0010(0.1180)$\\
 & \textcolor{blue}{$0.1200\pm 0.0012(0.1202)$} & \textcolor{blue}{$0.1169\pm 0.0010(0.1173)$} 
 & \textcolor{blue}{$0.1175\pm 0.0010(0.1174)$} & \textcolor{blue}{$0.1154\pm 0.0009(0.1151)$}  & \textcolor{blue}{$0.1149\pm 0.0008(0.1143) $}  \\

{\boldmath$100\theta{}_{s }$} & $1.04189\pm 0.00029 (1.04207)$
& $1.04199\pm 0.00030(1.04194)$& $1.04196\pm 0.00029(1.04181) $& $1.04197\pm 0.00029(1.04167)$ & $1.04199\pm 0.00031(1.04168)$\\

  & \textcolor{blue}{$1.04190^{+0.00027}_{-0.00031} (1.04178)$} &\textcolor{blue}{$1.04218\pm 0.00028(1.04211)$}     &  \textcolor{blue}{$1.04213\pm 0.00027(1.04225)$}  & \textcolor{blue}{$1.04236\pm 0.00028(1.04242)$} & \textcolor{blue}{$1.04242\pm 0.00029(1.04218) $} \\

{\boldmath$ln(10^{10}A_{s })$}& $3.040\pm 0.014 (3.046)  $& $3.039\pm 0.015(3.034)$ & $3.042\pm 0.014(3.044)$& $3.039\pm 0.014(3.038)$
& $3.037\pm 0.014(3.045) $\\
& \textcolor{blue}{$3.046\pm 0.014(3.049)$}  &\textcolor{blue}{$3.058^{+0.014}_{-0.017}(3.053)$}     &  \textcolor{blue}{$3.056\pm 0.016(3.047)$}  & \textcolor{blue}{$3.064^{+0.015}_{-0.017}(3.063)$}  & \textcolor{blue}{$3.062^{+0.013}_{-0.016} (3.079)$} \\

{\boldmath$n_{s }         $}& $0.9669\pm 0.0043(0.9664)$ & $0.9695\pm 0.0041(0.9692)$& $0.9679\pm 0.0039(0.9644)$ & $0.9682\pm 0.0040( 0.9711)$ & $0.9695\pm 0.0043( 0.9701)$\\
& \textcolor{blue}{$0.9657\pm 0.0041(0.9658) $}  &\textcolor{blue}{$0.9733\pm 0.0039(0.9706)$}    & \textcolor{blue}{$0.9715\pm 0.0035(0.9728) $ } & \textcolor{blue}{$0.9768\pm 0.0038(0.9801) $ }& \textcolor{blue}{$0.9786\pm 0.0035(0.9797)$} \\

{\boldmath$\tau{}_{reio } $}& $0.0528\pm 0.0073 (0.0569) $& $0.0532\pm 0.0077(0.0515)$ & $0.0534\pm 0.0073(0.0544)$& $0.0522\pm 0.0073 ( 0.0555)$& $0.0525\pm 0.0074(0.0584)$\\
& \textcolor{blue}{$0.0550\pm 0.0072(0.5488)$}  &\textcolor{blue}{$0.0639^{+0.0073}_{-0.0087}(0.0608)$}     &\textcolor{blue}{ $0.0624^{+0.0074}_{-0.0086}(0.0586)$}  & \textcolor{blue}{$0.0684^{+0.0076}_{-0.0089}(0.0685)$}  & \textcolor{blue}{$0.0678^{+0.0067}_{-0.0085}(0.0771)$} \\

{\boldmath$z_{\dagger}             $}& unconstrained & $1.70^{+0.09}_{-0.19}(1.65)$& $1.87^{+0.13}_{-0.21}(1.75)$ & $1.70^{+0.10}_{-0.13}(1.67) $ & $1.72^{+0.09}_{-0.12}(1.70)$\\

& \textcolor{blue}{$--$} &\textcolor{blue}{$--$}      & \textcolor{blue}{$--$}  & \textcolor{blue}{$--$} & \textcolor{blue}{$--$}\\

\hline


{\boldmath$z_{reio }      $}& $7.43^{+0.78}_{-0.67}(7.83) $ & $7.42\pm 0.78(7.25)$ & $7.47\pm 0.74( 7.59)$& $7.34\pm 0.76(7.67)$ & $7.34\pm 0.74(7.94) $\\
& \textcolor{blue}{$7.75\pm 0.72(7.76)$} &\textcolor{blue}{$8.52\pm 0.76(8.25)  $} & \textcolor{blue}{$8.39\pm 0.76(8.05)$}      & \textcolor{blue}{$8.87\pm 0.75(8.88) $}  & \textcolor{blue}{$8.79^{+0.64}_{-0.75}(9.64)$}\\

{\boldmath$Y_{\rm P}            $}& $0.247856\pm 0.000063 (0.247905)$& $0.247889\pm 0.000060( 0.247901)$& $0.247876\pm 0.000058 ( 0.247881) $ & $0.247877\pm 0.000061(0.247887)$& $0.247895\pm 0.000057(0.247903)$\\
& \textcolor{blue}{$0.247842\pm 0.000062(0.247832) $}  &\textcolor{blue}{$0.247944\pm 0.000059(0.247914)$}      & \textcolor{blue}{$0.247921^{+0.000059}_{-0.000053}(0.247888)$}  & \textcolor{blue}{$0.248010\pm 0.000056 ( 0.248020) $} & \textcolor{blue}{$0.248031\pm 0.000055 (0.248034)$} \\

{\boldmath$z_{d }         $}& $1060.03\pm 0.29(1060.22) $& $1060.15\pm 0.29(1060.22)$ & $1060.12\pm 0.29(1060.19)$ & $1060.12\pm 0.30(1060.15)$& $1060.15\pm 0.27 (1060.16)$\\
& \textcolor{blue}{$1059.99\pm 0.28(1059.95)$}  &\textcolor{blue}{$1060.28\pm 0.28(1060.16)  $}     & \textcolor{blue}{$1060.21\pm 0.27 (1060.03)$} & \textcolor{blue}{$1060.52\pm 0.28(1060.55) $}  & \textcolor{blue}{$1060.59\pm 0.29 (1060.56) $} \\

{\boldmath$r_{d }  {\rm[Mpc]}      $}& $147.17\pm 0.27 (147.28) $ & $147.31\pm 0.26(147.30)$& $147.20\pm 0.25(147.03) $& $147.21\pm 0.24(147.33)$ & $147.36\pm 0.23(147.46) $\\
& \textcolor{blue}{$147.07^{+0.24}_{-0.27}(147.06) $}  &\textcolor{blue}{$147.65\pm 0.25(147.63)$ }    & \textcolor{blue}{$147.55^{+0.24}_{-0.21}(147.65)$} & \textcolor{blue}{$147.87\pm 0.23(147.93) $}  & \textcolor{blue}{$147.96^{+0.25}_{-0.23}(148.10)$} \\

{\boldmath$t_0   {\rm[Gyr]}         $}& $13.620^{+0.120}_{-0.042} ( 13.596) $ & $13.517^{+0.038}_{-0.049} ( 13.502) $& $13.576^{+0.039}_{-0.034}(13.560)$& $13.531\pm 0.028(13.524)$& $13.522\pm 0.027(13.521)$\\
& \textcolor{blue}{$13.793\pm 0.023 (13.800)$}  &\textcolor{blue}{$13.745\pm 0.021(13.756) $}     & \textcolor{blue}{$13.755\pm 0.020(13.760)$}  & \textcolor{blue}{$13.716\pm 0.020 ( 13.710)  $}  & \textcolor{blue}{$13.706\pm 0.018(13.709) $} \\

{\boldmath$M_{\rm B}    {\rm[mag]}        $}& $--$&$ --$& $-19.317^{+0.021}_{-0.025}(-19.311)$& $-19.290\pm 0.017(-19.278)$ & $-19.282\pm 0.017 (-19.280)$\\
 & \textcolor{blue}{$--$ } & \textcolor{blue}{$--$ }      & \textcolor{blue}{$-19.407\pm 0.013(-19.411)$} & \textcolor{blue}{$-19.379\pm 0.012(-19.373)$} & \textcolor{blue}{$-19.372\pm 0.011(-19.369)$} \\

{\boldmath$H_0 {\rm[km/s/Mpc]}            $}& $70.77^{+0.79}_{-2.70}(71.22) $ & $73.30^{+1.20}_{-1.00}(73.59)$& $71.72^{+0.73}_{-0.92}(71.97) $ & $72.82\pm 0.65(73.20) $& $73.16\pm 0.64(73.36)$\\
 & \textcolor{blue}{$67.39\pm 0.55(67.28)$ }&\textcolor{blue}{$68.84\pm 0.48(68.61)$}       & \textcolor{blue}{$68.55\pm 0.44( 68.54)$} & \textcolor{blue}{$69.57\pm 0.42 ( 69.73) $} & \textcolor{blue}{$69.83\pm 0.37(69.96)$} \\

{\boldmath$\omega{}_{m }  $}& $0.1426\pm 0.0011 (0.1418) $& $0.1418\pm 0.0011(0.1418)$& $0.1423\pm 0.0010(0.1429)$ & $0.1422\pm 0.0010(0.1418)$& $0.1416\pm 0.0010(0.1411)$\\
 & \textcolor{blue}{$0.1431\pm 0.0011(0.1432)$}   &\textcolor{blue}{$0.1401\pm 0.0010(0.1405)$ }     & \textcolor{blue}{$0.1407\pm0.0010(0.1406)$}  & \textcolor{blue}{$0.1388\pm 0.0009(0.1385) $ }& \textcolor{blue}{$0.1384\pm 0.0008 (0.1378)$ }     \\

{\boldmath$\Omega{}_{m }  $}& $0.2860^{+0.0230}_{-0.0099}(0.2796)$& $0.2643^{+0.0072}_{-0.0090}(0.2618)$& $0.2768^{+0.0072}_{-0.0063}(0.2759)$ & $0.2683\pm 0.0052(0.2646)$ & $0.2646\pm 0.0052(0.2622)$\\ 
 & \textcolor{blue}{$0.3151\pm 0.0075(0.3163) $}  &\textcolor{blue}{$0.2958\pm 0.0061(0.2984)$}      & \textcolor{blue}{$0.2995\pm 0.0056(0.2992)$} & \textcolor{blue}{$0.2869\pm 0.0051(0.2849) $}& \textcolor{blue}{$0.2837\pm 0.0045 (0.2816) $}    \\

{\boldmath$\sigma_8         $}& $0.8210^{+0.0064}_{-0.0110}(0.8191)$& $0.8278\pm 0.0086(0.8260)$& $0.8240\pm 0.0074(0.8281)$ & $0.8277\pm 0.0075(0.8274)$& $0.8244\pm 0.0067(0.8264)$\\
& \textcolor{blue}{$0.8121^{+0.0055}_{-0.0061}(0.8136) $}  &\textcolor{blue}{$0.8076^{+0.0058}_{-0.0067}( 0.8064)$ }     & \textcolor{blue}{$0.8087\pm 0.0062(0.8054)$}  & \textcolor{blue}{$0.8054\pm 0.0064(0.8047)$} & \textcolor{blue}{$0.8030\pm 0.0055(0.8076)$} \\

{\boldmath$S_8             $}& $0.801^{+0.026}_{-0.016} (0.791)$& $0.777\pm 0.011(0.772)$& $0.791\pm 0.011(0.794)
$ & $0.783\pm 0.010(0.777)$& $0.774\pm 0.009(0.773)(0.773)$\\
& \textcolor{blue}{$0.832\pm 0.013(0.835)$}  &\textcolor{blue}{$0.802\pm 0.011(0.804) $}      & \textcolor{blue}{$0.808\pm 0.010(0.804)$} & \textcolor{blue}{$0.788\pm 0.010(0.784) $}  & \textcolor{blue}{$0.781\pm 0.008 (0.782)$} \\

{\boldmath$D_H(2.33)/r_{\rm d}$}& $8.960^{+0.280}_{-0.380}(9.218)$& $9.240^{+0.035}_{-0.025}(9.252)$& $9.201^{+0.041}_{-0.017}(9.222)$ & $9.232\pm 0.025(9.242)$& $9.249\pm 0.025(9.261) $\\
& \textcolor{blue}{$8.615\pm 0.013(8.614)$}  & \textcolor{blue}{$8.648\pm 0.011(8.643)$}      & \textcolor{blue}{$8.641\pm 0.010(8.639)$}  & \textcolor{blue}{$8.664\pm 0.008( 8.667)$} & \textcolor{blue}{$8.670\pm 0.008(8.675) $} \\

\hline


 {\boldmath$\chi^2_{\rm min}$}& $2778.06$& $2793.38$& $4219.68$& $4097.32$& $4185.34$\\
 & \textcolor{blue}{$2780.52$}  & \textcolor{blue}{$2820.30$}  & \textcolor{blue}{$4235.18$} & \textcolor{blue}{$4138.26$}  & \textcolor{blue}{$4226.50$} \\

{\boldmath$\rm{ln} \mathcal{Z}$} & $-1423.17$& $-1432.71$& $-2144.75$& $-2084.37$& $-2133.85$\\
& \textcolor{blue}{$-1424.45$}& \textcolor{blue}{$-1445.36$}    & \textcolor{blue}{$-2152.27$}  & \textcolor{blue}{$-2103.84$}  & \textcolor{blue}{$-2153.62$} \\

{\boldmath${\rm ln} \mathcal{B}_{ij}$}& 
 $-1.28$   &$-12.65$  & $-7.52$ & $-19.47$ & $-19.77$ \\
 \hline
 \hline
\end{tabular}
}
\end{table*}

\begin{table*}[t!]

\caption{Concordance/discordance between the $\Lambda$CDM/$\Lambda_{\rm s}$CDM models and the theoretical/direct observational estimations, viz., $H_0=73.04\pm1.04~{\rm km\, s^{-1}\, Mpc^{-1}}$ (SH0ES)~\cite{Riess:2021jrx}; $M_{\rm B}=-19.244\pm 0.037\,\rm mag$ (SH0ES)~\cite{Camarena:2021dcvm}; $D_H(2.33)/r_{\rm d}=8.99\pm0.19$ (for the combined Ly-$\alpha$ data)~\cite{duMasdesBourboux:2020pck}; $t_{0}=13.50\pm0.15\,\rm Gyr$ (systematic uncertainties are not included)~\cite{Valcin:2021jcg}; $10^2\omega_{\rm b}^{\rm LUNA}=2.233\pm0.036$ (empirical approach, based primarily on experimentally measured cross sections for $d(p,\gamma)^3\rm He$ reaction)~\cite{Mossa:2020gjc} and $10^2\omega_{\rm b}^{\rm PCUV21}=2.195\pm0.022$ (theoretical approach, incorporating nuclear theory for $d(p,\gamma)^3\rm He$ reaction)~\cite{Pitrou:2020etk}. $S_8 = 0.746^{+0.026}_{-0.021}  $ ($\Lambda_{\rm s}$CDM: KiDS-1000) and $S_8 =0.749^{+0.027}_{-0.020}$ ($\Lambda$CDM: KiDS-1000) obtained in this work.}
\label{tab:tensions}
     \scalebox{0.72}{
 \begin{tabular}{lccccc}
  	\hline
    \toprule
    \textbf{Data set } & \textbf{Planck}& \textbf{Planck+BAOtr} & \textbf{Planck+BAOtr} \;\; & \textbf{Planck+BAOtr}\;\; & \textbf{Planck+BAOtr}   \\
 &  & & \textbf{+PP} \;\; & \textbf{+PP\&SH0ES}\;\; & \textbf{+PP\&SH0ES+KiDS-1000}   \\  \hline
      \textbf{Model} & \textbf{$\bm{\Lambda}_{\textbf{s}}$CDM}&
       \textbf{$\bm{\Lambda}_{\textbf{s}}$CDM}&\textbf{$\bm{\Lambda}_{\textbf{s}}$CDM}&\textbf{$\bm{\Lambda}_{\textbf{s}}$CDM}&\textbf{$\bm{\Lambda}_{\textbf{s}}$CDM}\\
        & \textcolor{blue}{\textbf{$\bm{\Lambda}$CDM}} & \textcolor{blue}{\textbf{$\bm{\Lambda}$CDM}} & \textcolor{blue}{\textbf{$\bm{\Lambda}$CDM}} & \textcolor{blue}{\textbf{$\bm{\Lambda}$CDM}} & \textcolor{blue}{\textbf{$\bm{\Lambda}$CDM}} 
          \\ \hline
      \vspace{0.1cm}
{\boldmath$H_0$}  & $1.4\sigma$ & $0.2\sigma$  & $1.0\sigma$  & $0.2\sigma$  & $0.1\sigma$ \\& \textcolor{blue}{$4.8\sigma$} &  \textcolor{blue}{$3.7\sigma$} &  \textcolor{blue}{$4.3\sigma$} &  \textcolor{blue}{$3.1\sigma$} &  \textcolor{blue}{$2.9\sigma$}
\\
{\boldmath$M_{\rm B}$}  & -- &   --  & $1.7\sigma$ & $1.1\sigma$  & $0.9\sigma$ \\ &  \textcolor{blue}{--} &  \textcolor{blue}{--} &  \textcolor{blue}{$4.5\sigma$}  &  \textcolor{blue}{$3.5\sigma$} &  \textcolor{blue}{$3.3\sigma$}
\\
{\boldmath$S_{8}$}   & $1.7\sigma$ & $1.2\sigma$& $1.7\sigma$   & $1.4\sigma$   
& $1.1\sigma$\\
 &  \textcolor{blue}{$3.1\sigma$}  & \textcolor{blue}{$2.0\sigma$} &  \textcolor{blue}{$2.3\sigma$} &  \textcolor{blue}{$1.5\sigma$}  &  \textcolor{blue}{$1.3\sigma$}
\\
{\boldmath$t_{\rm 0}$}  & $1.0\sigma$ & $0.1\sigma$   & $0.5\sigma$   & $0.2\sigma$ 
& $0.1\sigma$  \\  &  \textcolor{blue}{$1.9\sigma$} &  \textcolor{blue}{$1.6\sigma$} &  \textcolor{blue}{$1.7\sigma$}  &  \textcolor{blue}{$1.4\sigma$} &  \textcolor{blue}{$1.4\sigma$}
\\
{\boldmath$D_H(2.33)/r_{\rm d}$}  & $0.2\sigma$ & $1.3 \sigma$ & $1.1 \sigma$    & $1.3 \sigma$ & $1.4 \sigma$  \\ &  \textcolor{blue}{$2.0\sigma$} &  \textcolor{blue}{$1.8 \sigma$} &  \textcolor{blue}{$1.8 \sigma$} &  \textcolor{blue}{$1.7 \sigma$} &  \textcolor{blue}{$1.7 \sigma$}
\\
{\boldmath$\omega_{\rm b }^{\rm PCUV21}$}   & $1.2\sigma$  & $2.1\sigma$   & $1.9\sigma$& $1.9\sigma$ & $2.2\sigma$  \\ &  \textcolor{blue}{$1.6\sigma$} &  \textcolor{blue}{$2.6\sigma$}  &  \textcolor{blue}{$2.4\sigma$}  &  \textcolor{blue}{$3.1\sigma$} &  \textcolor{blue}{$3.4\sigma$} 
\\
{\boldmath$\omega_{\rm b}^{\rm LUNA}$}  & $0.3\sigma$   & $0.4\sigma$  & $0.3\sigma$   & $0.3\sigma$ & $0.4\sigma$  \\ &  \textcolor{blue}{$0.1\sigma$} &  \textcolor{blue}{$0.8\sigma$} &  \textcolor{blue}{$0.6\sigma$}  &  \textcolor{blue}{$1.1\sigma$} &  \textcolor{blue}{$1.3\sigma$}
\\
\hline
 \hline
\end{tabular}
}
\end{table*}

\section{Triangle Posteriors}
\label{sec:Appendix}
\begin{figure*}[htbp]
    \centering
    \includegraphics[width=0.85\textwidth]{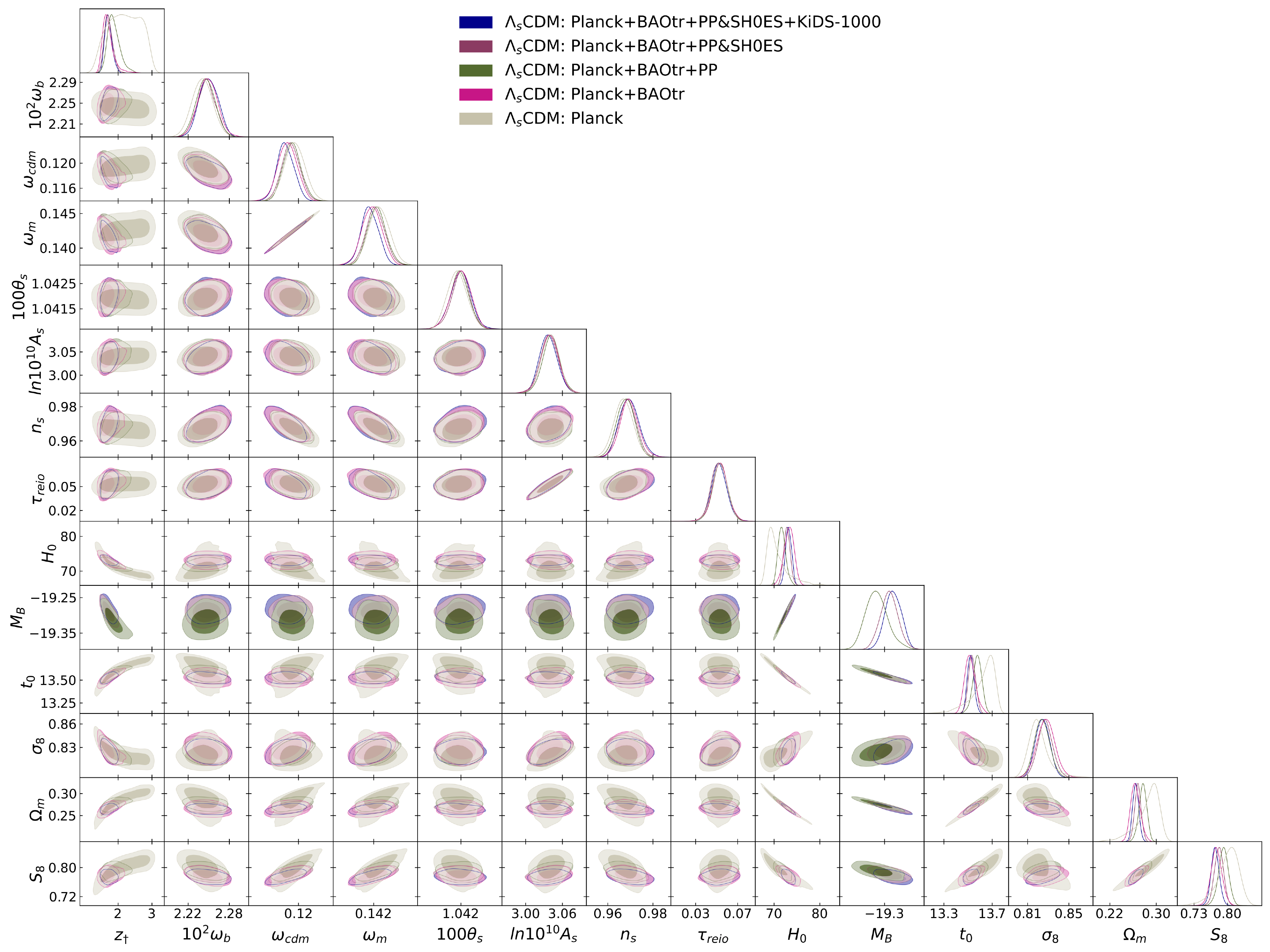}
    \caption{One- and two-dimensional (68\% and 95\% CLs) marginalized distributions of the $\Lambda_{\rm s}$CDM model parameters from Planck, Planck+BAOtr, Planck+BAOtr+PP, Planck+BAOtr+PP\&SH0ES, and Planck+BAOtr+PP\&SH0ES+KiDS-1000.}
    \label{fig:lscdmtc}
    \bigskip
    \includegraphics[width=0.85\textwidth]{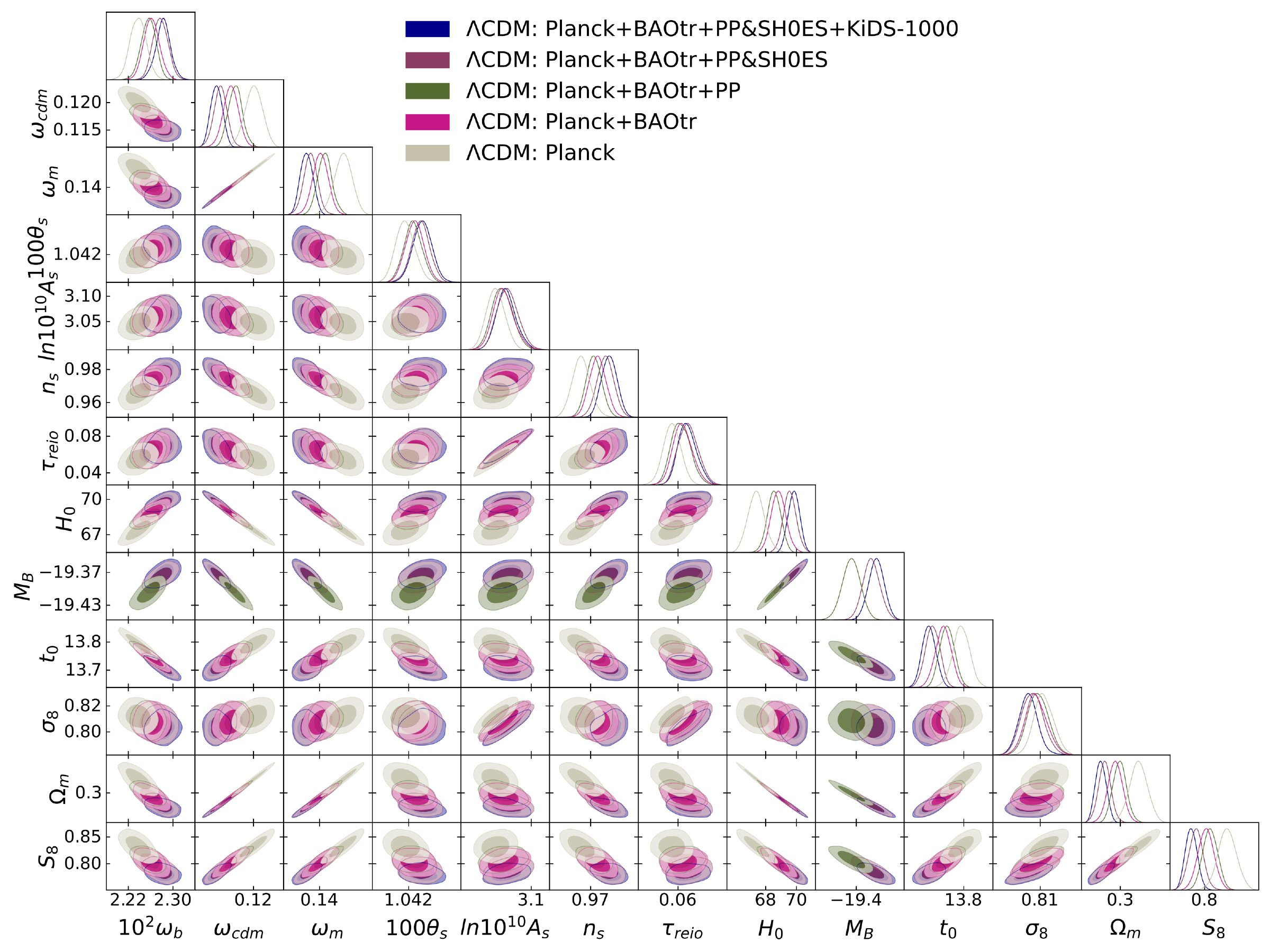}
    \caption{ One- and two-dimensional (68\% and 95\% CLs) marginalized distributions of the $\Lambda$CDM model parameters from Planck, Planck+BAOtr, Planck+BAOtr+PP, Planck+BAOtr+PP\&SH0ES, and Planck+BAOtr+PP\&SH0ES+KiDS-1000.}
    \label{fig:lcdmtc}
\end{figure*}

\begin{figure*}[htbp]
   \includegraphics[width=0.85\textwidth]{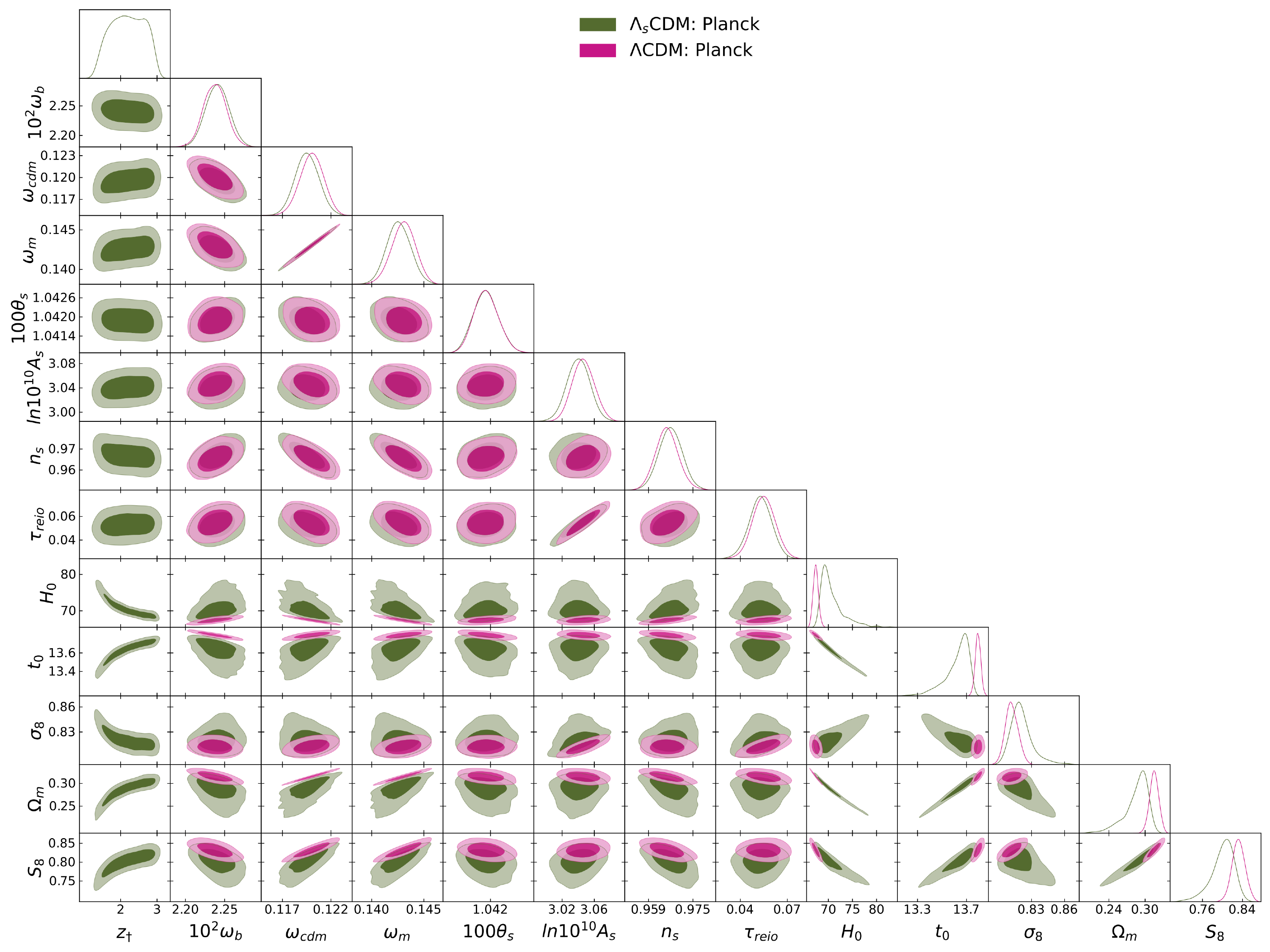}     \label{fig:Planck}
    \caption{ One- and two-dimensional (68\% and 95\% CLs) marginalized distributions of the model parameters from Planck.}
    \bigskip
    \includegraphics[width=0.85\textwidth]{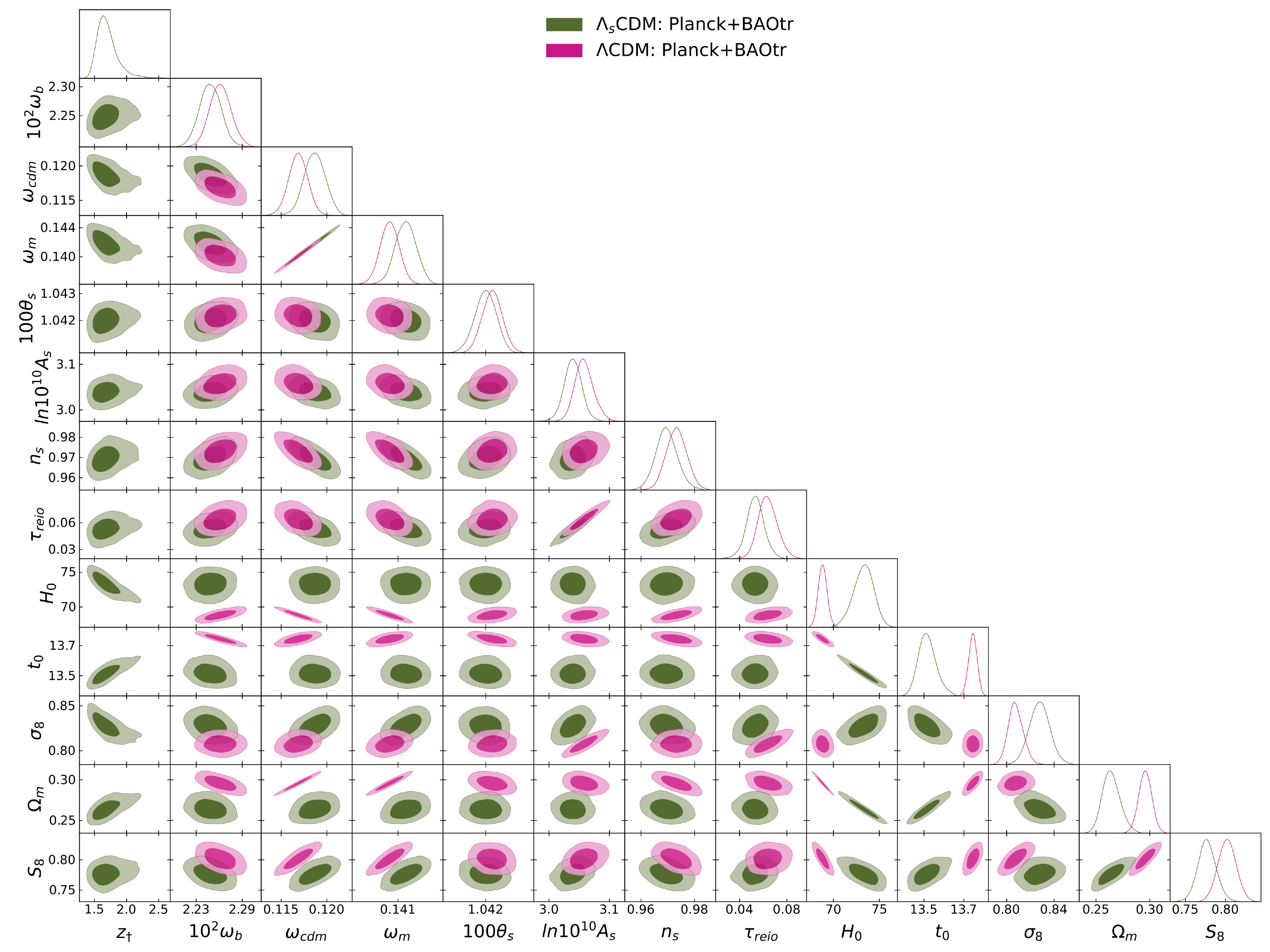}
    \caption{ One- and two-dimensional (68\% and 95\% CLs) marginalized distributions of the model parameters from Planck+BAOtr.}
    \label{fig:Pbtr}
   \end{figure*}

\begin{figure*}
    \centering
    \includegraphics[width=0.85\textwidth]{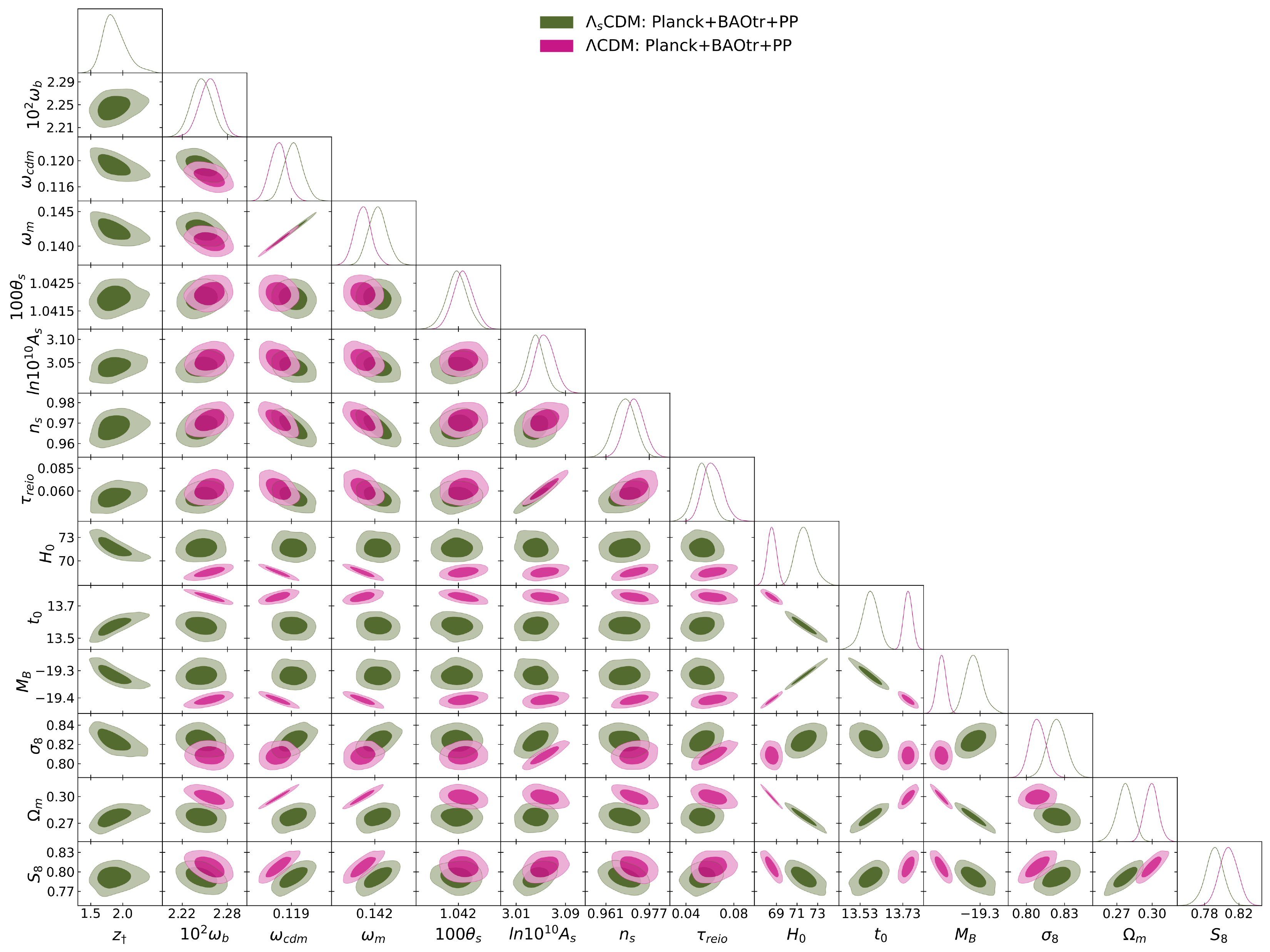}
    \caption{ One- and two-dimensional (68\% and 95\% CLs) marginalized distributions of the model parameters from  Planck+BAOtr+PP.}
    \label{fig:PbtrPP}
    \bigskip
    \includegraphics[width=0.85\textwidth]{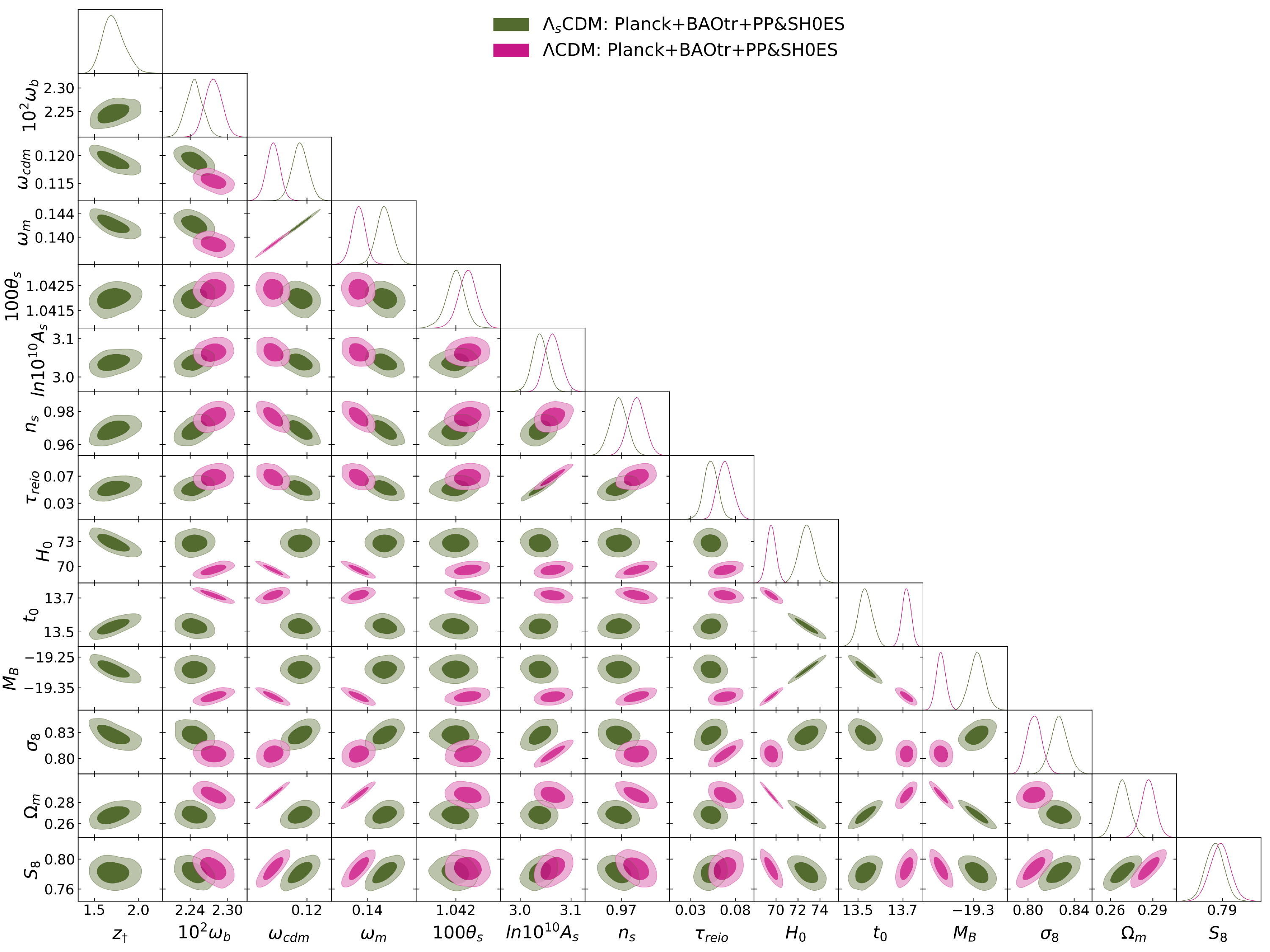}
    \caption{ One- and two-dimensional (68\% and 95\% CLs) marginalized distributions of the model parameters from  Planck+BAOtr+PP\&SH0ES.}
    \label{fig:PbtrPPS}
    \end{figure*}

\begin{figure*}
    \centering
    \includegraphics[width=0.85\textwidth]{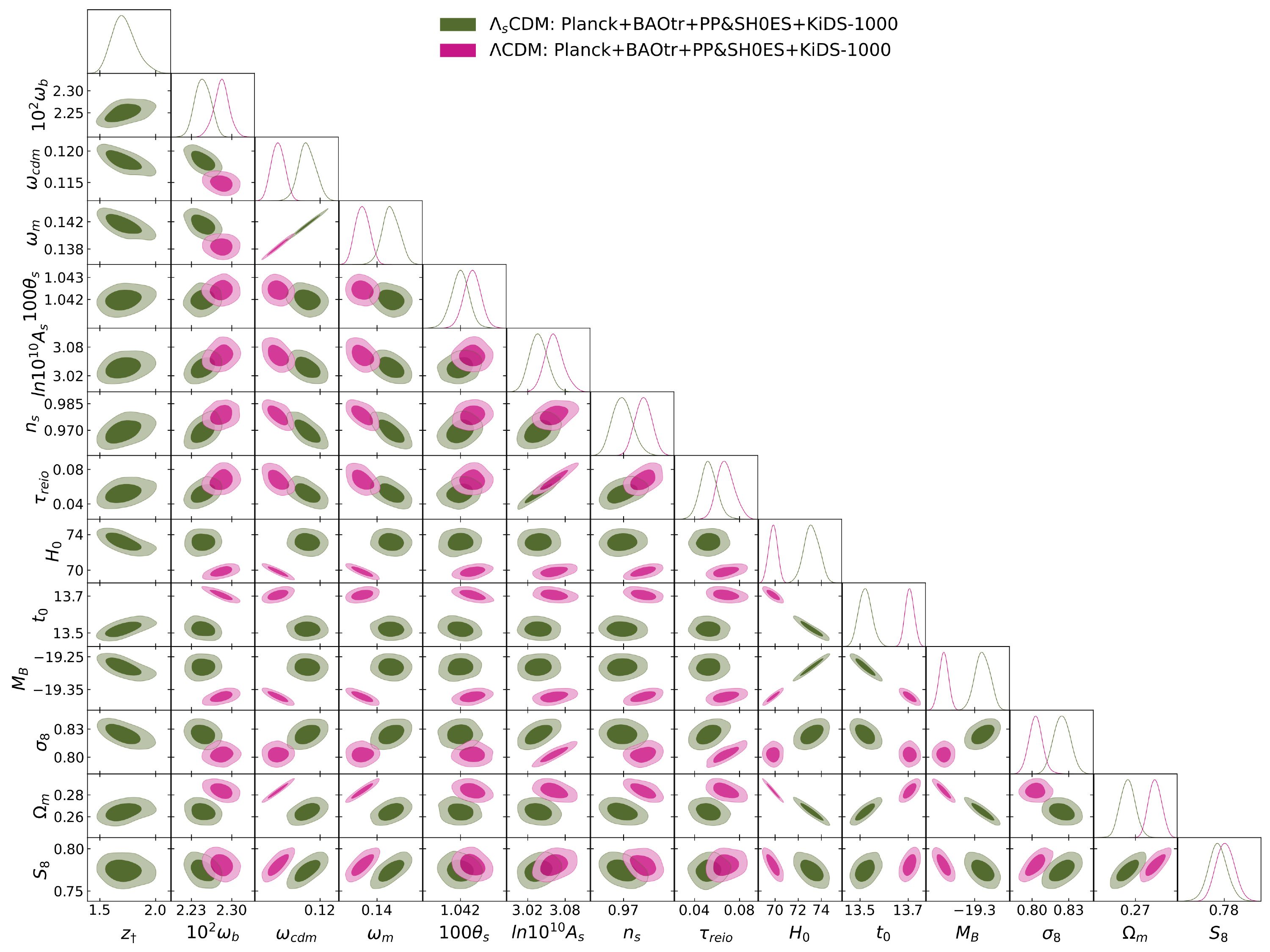}
    \caption{ One- and two-dimensional (68\% and 95\% CLs) marginalized distributions of the model parameters from  Planck+BAOtr+PP \&SH0ES+KiDS-1000.}
    \label{fig:PbtrPPSK}
\end{figure*}

In Fig.~\ref{fig:lscdmtc}-\ref{fig:PbtrPPSK}, we present the
One- and two-dimensional (at 68\% and 95\% CL) marginalized distributions of the model parameters for both models. We do not see strong correlations between $z_\dagger$ and the six baseline parameters, but these exist among $z_\dagger,\,H_0,\,M_{\rm B},\,S_8$, and $\Omega_{\rm m}$. Thus,  triangular plots showing the joint posteriors between the parameters present extra complementary information to the tables in the main text.

\bibliography{supp}